\documentclass[final,1p,times]{elsarticle}
\biboptions{sort&compress}

\usepackage{lineno}
\usepackage{amsmath}
\usepackage{amssymb}
\usepackage{aas_macros}
\usepackage{bm}
\usepackage{soul}

\usepackage{graphicx}
\usepackage{epsfig}
\usepackage{epstopdf}
\usepackage{color}
\usepackage{xcolor}
\usepackage{colortbl}
\usepackage{multirow}
\usepackage{mleftright}

\usepackage{url}
\usepackage{xspace}
\usepackage{comment}
\usepackage{braket}
\usepackage{hhline}

\usepackage{hyperref}
\usepackage{cleveref}

\newcommand{\kpar}{k_{\parallel}}
\newcommand{\kparmin}{k_{\parallel, {\rm min}}}
\newcommand{\kperp}{k_{\bot}}

\newcommand{\bk}{\bm{k}}
\newcommand{\bq}{\bm{q}}
\newcommand{\bx}{\bm{x}}
\newcommand{\fnl}{f_{\rm NL}}
\newcommand{\fnll}{f_{\mathrm{NL}}^{\mathrm{loc}}}
\newcommand{\fnle}{f_{\mathrm{NL}}^{\mathrm{equil}}}
\newcommand{\fnlo}{f_{\mathrm{NL}}^{\mathrm{orth}}}
\newcommand{\Mpc}{\ensuremath{\text{$h$/Mpc}}\xspace}
\newcommand*\mean[1]{\overline{#1}}

\newcommand*{\ie} {i.\,\!e.,\,}

\newcommand*{\veps} {\varepsilon}
\newcommand*{\eref} [1] {Eq.\ \eqref{#1}}
\newcommand*{\esref} [2] {Eqs.\ \eqref{#1}\ and\ \eqref{#2}}

\newcommand*{\tref} [1] {Table \ref{#1}\xspace}
\newcommand*{\sref} [1] {Sec.\ \ref{#1}\xspace}
\newcommand*{\fref} [1] {Fig. \ref{#1}\xspace}

\newcommand*{\perm} {\;\text{perm}}

\newcommand{\rev}[1]{{#1}}

\graphicspath{{./plots/}}

\journal{Physics of the Dark Universe}

\begin{document}

\begin{frontmatter}




\title{{Probing primordial non-Gaussianity with the \rev{power spectrum and} bispectrum of future 21cm intensity maps}}

\author[1]{Dionysios Karagiannis\corref{cor1}}
\ead{dakaragian@gmail.com}
\cortext[cor1]{Corresponding author.}
\author[2,3,1,8]{Jos\'e Fonseca}
\author[1,4]{Roy Maartens}
\author[5,6,7,1]{Stefano Camera}

% List of institutions
\address[1]{Department of Physics \& Astronomy, University of the Western Cape,\\Cape Town 7535, South Africa}
\address[2]{Dipartimento di Fisica e Astronomia ``G. Galilei'', Universit\`a degli Studi di Padova,\\Via Marzolo 8, 35131 Padova, Italy}
\address[3]{INFN -- Istituto Nazionale di Fisica Nucleare, Sezione di Padova,\\Via Marzolo 8, 35131 Padova, Italy}
\address[4]{Institute of Cosmology \& Gravitation, University of Portsmouth, Portsmouth PO1 3FX, UK}
\address[5]{Dipartimento di Fisica, Universit\`a degli Studi di Torino,\\Via P.\ Giuria 1, 10125 Torino, Italy}
\address[6]{INFN -- Istituto Nazionale di Fisica Nucleare, Sezione di Torino,\\Via P.\ Giuria 1, 10125 Torino, Italy}
\address[7]{INAF -- Istituto Nazionale di Astrofisica, Osservatorio Astrofisico di Torino,\\Strada Osservatorio 20, 10025 Pino Torinese, Italy}
\address[8]{School of Physics \& Astronomy, Queen Mary University of London, London E1 4NS, UK}

\begin{abstract}
After reionisation, the 21cm emission line of neutral hydrogen within galaxies provides a tracer of dark matter. Next-generation intensity mapping surveys, with the SKA and other radio telescopes, will cover large sky areas and a wide range of redshifts, facilitating their use as probes of primordial non-Gaussianity. {Previous works have shown that the bispectrum can achieve tight constraints on primordial non-Gaussianity with future surveys that are purposely designed for intensity mapping in interferometer mode}. Here we investigate the constraints attainable from surveys operating in single-dish mode, \rev{using the combined power spectrum and bispectrum signal}. In the case of the power spectrum, single-dish surveys typically outperform interferometer surveys. We find that the reverse holds for the bispectrum: single-dish surveys are not competitive with surveys designed for interferometer mode.
\end{abstract}


\end{frontmatter}

% \linenumbers

\section{Introduction}

{Observations of the cosmic microwave background (CMB) and studies of its anisotropies in temperature and polarisation \cite{2020A&A...641A...6P,2020A&A...641A...1P} have confirmed to a high degree of accuracy our current description of the (early) Universe in terms of the concordance $\Lambda$CDM cosmological model. On the other end of the spectrum, low-redshift measurements of the cosmic large-scale structure (LSS) point towards the same picture \cite{2020arXiv200709008D,2020arXiv200709007R,2019PhRvL.122q1301A,2020arXiv200701844J,2019PhRvD.100b3541A}. Nonetheless, several major questions remain unsolved, like the mechanism that drove the cosmological inflationary period in the primordial Universe, responsible for the formation of the  seeds of both the CMB anisotropies and the LSS.}

{Inflation is the umbrella term for a family of theories describing how quantum fluctuations in the primordial Universe  evolved to a macroscopic level, thus becoming the seeds of cosmic structures. One of the most common predictions of inflation---the so-called `smoking guns'---is the presence of a certain (tiny) amount of non-Gaussianity in the distribution of primordial density perturbations. It is useful to parametrise such a primordial non-Gaussianity (PNG) in terms of $\fnl$, namely the amplitude of the first term in a Taylor expansion around Gaussianity. Measurements of, or bounds on, this parameter have the potential to rule out entire classes of inflationary models, thus strengthening our understanding of the early phases of the Universe's evolution.}

{Currently, the tightest constraints on $\fnl$ come from bounds on the amplitude of the bispectrum of CMB anisotropies \cite{1905.05697}, which for instance constrain so-called local-type PNG to be $\fnll=-0.9\pm5.1$ at $68\%$ CL\ (more details on different types of PNG are given in the next section). However, most of the information on PNG has already been extracted from the CMB, and the next frontier is surveys of the LSS, which provide two complementary probes: the bispectrum  (e.g. \cite{Sefusatti:2007ih}) and the scale-dependent power spectrum  of biased tracers (e.g. \cite{2008ApJ...684L...1C}). The latter has already been investigated with catalogues from state-of-the-art galaxy surveys, and has provided  complementary constraints on $\fnl$ (e.g. \cite{Giannantonio:2013uqa,Karagiannis:2013xea,Castorina:2019wmr}). In this paper, we focus  \rev{instead on the combined power spectrum and bispectrum signal}, with a new angle offered by forthcoming cosmological experiments at radio frequencies.}

{Cosmology in the radio band traditionally offered two main probes, both based on the study of galaxy clustering: continuum galaxies (e.g. \cite{Blake:2001bg,Overzier:2003kg}) and neutral hydrogen (HI) 21cm emission-line galaxies (e.g. \cite{2012ApJ...750...38M,Papastergis:2013lia}). 
% The former has the advantage of detecting a large number of galaxies over almost the full sky. On the one hand, this has allowed for detailed angular clustering analyses [NVSS, the other beginning with `T' by Enzo Branchini, possibly others], but continuum measurements lack redshift information, thus precluding the possibility of studying the redshift evolution of cosmic structures. The latter, on the other hand, comes with precise redshift estimates, directly from the width of the 21cm line, but it generally targets small sky areas and source number density is small compared to optical counterparts, thus limiting to the study of the local Universe.
Each has its own advantages and disadvantages, but in this paper, we instead focus on a third probe proposed for cosmological studies: HI intensity mapping \cite{Bharadwaj:2000av,Battye:2004re,Wyithe2007,2008PhRvL.100i1303C,Battye2013}. In the post-reionisation Universe, most HI resides in dense systems inside galaxies and  thus provides us with a tracer of the cosmic LSS. {The HI intensity mapping technique consists of making maps of the brightness temperature of the sky at different frequencies. Since no other emission lines appear at these radio frequencies, there is a unique relation between observed frequency and redshift,}  $(1+z)\nu=\nu_{\rm HI}=c/\lambda_{\rm HI}$, with $\lambda_{\rm HI}=21\,\mathrm{cm}$ the rest-frame wavelength of the HI hyperfine transition photon. {Each pixel in the map contains many galaxies so that their combined emission yields a larger detectable signal. Finally, the temperature maps are analysed via summary statistics such as Fourier- or harmonic-space power spectra and bispectra.}}

The power spectrum of HI intensity mapping has already been suggested as a powerful probe to study PNG \cite{Camera2013,Xu2014,Fonseca:2015laa,2017MNRAS.466.2780F, Ballardini:2019wxj}, and it has been shown that single-dish mode is the best experimental set up for this specific goal. On the other hand, \cite{Karagiannis:2019jjx} has explored the potential of bispectrum measurements from future HI intensity mapping experiments in interferometer mode, {finding very competitive forecast results on PNG (e.g., $\sigma(\fnll)<1$ and $\sigma(\fnle)<5$)}. Here, we compare the capabilities of single-dish mode surveys with the interferometer mode results, \rev{while using the combined power spectrum and bispectrum signal}.
%, to have a better grasp on the physics behind such measurements {and conclude on an ideal experimental setup for a HI bispectrum analysis, and in particular, one that focuses on PNG}.}

% \roy{A comprehensive and realistic treatment of the problem would be to simulate the data, including foregrounds, perform foreground subtraction and do a joint power spectrum and bispectrum analysis. However, this is a major project which requires considerable further work (see e.g. \cite{Bernal:2019jdo,Bernal:2020lkd} for some recent work). Our aim is much more limited, i.e., to compare the bispectrum PNG constraints using single-dish as opposed to interferometer surveys. In order to do this, we use the same simplified Fisher analysis as in the interferometer case \cite{Karagiannis:2019jjx}. Since the power spectrum PNG constraints for single-dish surveys are known to outperform those of interferometer surveys, we confine our analysis to the bispectrum constraints alone.}

A comprehensive and realistic treatment of the problem would be to simulate the data, including foregrounds {and} performing foreground subtraction.  However, this is a major project which requires considerable further work (see e.g. \cite{Bernal:2019jdo,Bernal:2020lkd} for some recent analyses).  Our aim is more limited, focusing on the comparison of \rev{the joint power spectrum and} bispectrum PNG constraints using single-dish as opposed to interferometer surveys.  For the power spectrum, PNG constraints for single-dish surveys are known to outperform those of interferometer surveys.  However, this has not been assessed in the case of the  \rev{joint power spectrum and bispectrum signal}, and this is indeed the scope of our paper.  In order to do this, we use the same simplified Fisher analysis as in the interferometer case \cite{Karagiannis:2019jjx}.

{This paper is organised as follows. In \sref{sec:matter_PK_BK} we review the matter \rev{power spectrum and} bispectrum model, {as well as} the PNG types considered here. In \sref{sec:halo_bias} we present the formalism for the HI bias, while in \sref{sec:HIbisp_RSD} the final model for the \rev{power spectrum and} bispectrum of the HI fluctuations in redshift space is shown. In \sref{sec:surveys} the specifications of experiments under consideration  are presented. In \sref{sec:fisher} we review the Fisher matrix formalism used to forecast the amplitude of PNG, while in \sref{sec:obs_window} we discuss the observational limitations for each experimental mode assumed here. Finally, the results are presented in \sref{sec:results}, followed by a discussion in \sref{sec:discussion}.}

\section{Matter \rev{power spectrum and} bispectrum}\label{sec:matter_PK_BK}

The power spectrum of the Bardeen gauge-invariant primordial gravitational potential is defined in Fourier space by
\begin{equation}
 {\langle \Phi(\bk)\Phi(\bk')\rangle}=(2\pi)^3 \delta_{\rm D}(\bk+\bk')P_\Phi(k),  
\end{equation}
where $P_\Phi(k)$ is directly related to the power spectrum of the primordial curvature perturbations $\zeta$ (during the matter-dominated era, $\Phi(\bk)=3\zeta(\bk)/5$), which are generated during inflation. They are expected to have a nearly perfect Gaussian distribution in the case of the standard single-field slow-roll inflationary scenario, which means that they can be adequately characterised by their power spectrum. The primordial perturbations $\Phi$ are in turn related to the linear dark matter over-density field through the Poisson equation, $\delta_m^{\rm L}(\bk,z)=M(k,z)\Phi(\bk)$, where
 \begin{eqnarray}\label{eq:powlin}
 P_m^{\rm L}(k,z)&=&M^2(k,z)\,P_\Phi(k),\\
 \label{eq:poisM}
 M(k,z)&=&\frac{2c^2D(z)}{3\Omega_m H_0^2\, {g_{\rm dec}}}\,T(k)\, k^2.
 \end{eqnarray}
 Here $D(z)$ is the linear growth factor (since the linear fluid equations generate a linearly evolved matter density field), normalised to unity today (\ie $D(0)=1$), and $T(k)$ is the matter transfer function normalized to unity at large scales $k \rightarrow 0$. {The factor $g_{\rm dec}$, \ie the Bardeen potential growth factor at decoupling, ensures that $\fnl$ is in the CMB convention \cite{Camera:2014bwa,Desjacques2016}.}
 The linear power spectrum is computed with the numerical Boltzmann code CAMB \cite{CAMB}. 
 
 Deviations from the standard inflationary model will generate a violation of Gaussian initial conditions \cite{Komatsu2009}. The presence of PNG will generate nonzero higher-order correlators, where the most important is the bispectrum, i.e. the Fourier transform of the three-point function:
 \begin{equation}
  \langle\Phi(\bk_1)\Phi(\bk_2)\Phi(\bk_3)\rangle=(2\pi)^3\delta_{D}(\bk_1+\bk_2+\bk_3)B_{\Phi}(k_1,k_2,k_3).
 \end{equation}
 Here the Dirac delta function ensures the conservation of momentum and imposes a triangle condition, correlating fluctuations at three points in Fourier space\footnote{We use the ordering $k_3\le k_2\le k_1$.}. Although the number of shapes of the triangles can be large, violating different inflationary conditions generates bispectrum signals that peak at distinct triangle configurations. The strength of this PNG signal in the bispectrum is defined by a dimensionless parameter, $\fnl$. Thus, measuring the primordial bispectrum, and most importantly its amplitude, from a cosmological data-set, offers a unique opportunity to shed light on the initial phases of the Universe.  

 In this work, we will consider the most studied PNG shapes:
 \begin{itemize}
 \item {\em local} \cite{Salopek1990,Gangui1993,Verde1999,Komatsu2001}, which peaks for squeezed triangles ($k_3\ll k_2\simeq k_1$); 
 \item {\em equilateral} \cite{Creminelli2005}, peaking for equilateral configurations ($k_3\simeq k_2\simeq k_1$); 
 \item {\em orthogonal} \cite{Senatore2009}, which peaks for both equilateral and folded triangles ($k_1\simeq k_2 \simeq k_3/2$).
 \end{itemize}
 Theses templates are defined respectively as:
  \begin{align}
B_{\Phi}^{\text{loc}}(k_1,k_2,k_3)&=2 \fnll\Big[P_{\Phi}(k_1)P_{\Phi}(k_2)+\text{2 perms} \Big] \, ,\\
B_{\Phi}^{\text{eq}}(k_1,k_2,k_3)&=6 \fnle\bigg\{-\Big[P_{\Phi}(k_1)P_{\Phi}(k_2)+\text{2 perms} \Big] -2\Big[P_{\Phi}(k_1)P_{\Phi}(k_2)P_{\Phi}(k_3)\Big]^{2/3} \nonumber \\
&+\Big[P_{\Phi}^{1/3}(k_1)P_{\Phi}^{2/3}(k_2)P_{\Phi}(k_3)+\text{5 perms}\Big] \bigg\} \,, \\ 
 B_\Phi^\text{orth}(k_1,k_2,k_3) &= 6\fnlo\bigg\{ 3\Big[P_\Phi^{1/3}(k_1)P_\Phi^{2/3}(k_2)P_\Phi(k_3) +5\text{ perms}\Big] \nonumber \\
 &-3 \Big[P_{\Phi}(k_1)P_{\Phi}(k_2)+\text{2 perms} \Big]-8 \Big[P_\Phi(k_1)P_\Phi(k_2)P_\Phi(k_3)\Big]^{2/3}
\bigg\} \,.
\end{align} 
 Following the relation between the linear matter density contrast and the primordial potential, we find the leading order PNG contribution to the matter density bispectrum:
 \begin{equation} \label{eq:bisng}
  B_I(k_1,k_2,k_3,z)=M(k_1,z)M(k_2,z)M(k_3,z)B_\Phi(k_1,k_2,k_3) \,.
 \end{equation}
The matter bispectrum has additional terms besides the PNG contribution of \eref{eq:bisng}, due to the nonlinearities induced by gravity, even at zeroth order. Therefore, in order to retrieve the PNG information from the LSS bispectrum, we need good knowledge of the gravitational part of the bispectrum. Throughout this work, we will describe the nonlinearities in the framework of Standard Perturbation Theory (SPT) (see e.g. \cite{Bernardeau2002} for a review). 

\section{Bias of neutral hydrogen} \label{sec:halo_bias}

Forecasting the amplitude of PNG from the \rev{power spectrum and} bispectrum of future HI IM surveys (see \sref{sec:surveys} for details) requires a relation between the statistics of observed tracers and the underlying distribution of dark matter (see e.g.~\cite{Desjacques2016} for a review). The bias is a combination of two components: the bias relation between halos and dark matter and how the neutral hydrogen is distributed amongst the dark matter halos.  

Here we consider halo bias up to second order, which is sufficient for the spatial scales considered here, i.e. much larger scales than those involved in halo formation.
%In the past the bias expansion contained only powers of the dark matter contrast $\delta_m$ \cite{Coles1993,Fry1993,Fry1996,Catelan1997} (local-in-matter bias expansion), while later on they were supplemented by terms of the tidal field $s_{ij}$ \cite{Catelan2000,McDonald2009,Elia2010,Chan2012,Baldauf2012}. 
We use the approach of the {general bias expansion} \cite{Assassi2014,Senatore2014,Mirbabayi2014}, in which the halo over-density field $\delta_h$ is described as a function of all possible local gravitational observables, which are introduced in the expansion in the form of renormalised operators. A complete Eulerian bias expansion can be built from the tensor $\partial_i\partial_j\Phi$ and its convective time derivatives \cite{Mirbabayi2014}, where the first contains the trace $\nabla^2\Phi\propto\delta_m$ and the trace-free  tidal field $s_{ij}=\big(\partial_{i}\partial_{j}-\delta_{ij}\nabla^2/3\big)\nabla^{-2} \delta_m$.

For Gaussian initial conditions and up to second order, the Eulerian halo density contrast can be written as  \cite{Assassi2014,Senatore2014,Mirbabayi2014,Desjacques2016}:
  \begin{equation}\label{eq:deltaG}
   \delta_h^{E,G}(\bx,\tau)= b_1^E(\tau)\delta(\bx,\tau) +\veps^E(\bx,\tau)+\frac{b_2^E(\tau)}{2}\delta^2(\bx,\tau) + \frac{b_{s^2}^E(\tau)}{2}s^2(\bx,\tau)+\veps_{\delta}^E(\bx,\tau)\delta(\bx,\tau) \,,
  \end{equation}
  where $\tau$ is the conformal time, $\bx$ are the spatial comoving coordinates in the Eulerian frame, $s^2=s_{ij}s^{ij}$
  is the simplest scalar that can be formed from the tidal field,  
  %cannot appear at first order since its trace-free \footnote{At first order in the general bias expansion formalism and in the Eulerian framework, we have only $\rm{tr}[\Pi^{[1]}]$ \cite{Mirbabayi2014}, where $\Pi^{[1]}_{ij}(\bx,\tau)=s_{ij}(\bx,\tau)+\frac{1}{3}\delta_{ij}\delta(\bx,\tau)$, and therefore only $\delta_m$ can appear. } , 
  $\veps^E$ is the leading stochastic field \cite{Dekel1998,Taruya1998,Matsubara1999} and $\veps_{\delta}^E$ is the stochastic field associated with the linear bias. These fields take into account the stochastic relation between the galaxy density and any large-scale field. The second-order tidal field bias coefficient, following the convention in \cite{Baldauf2012}, is given by $b_{s^2}^E=-4(b_1^E-1)/7$.
  
  In the presence of PNG there is a scale-dependent correction to the linear bias $b_1^E$, especially in the case of local PNG \cite{Dalal2008,Slosar2008,Matarrese2008,Verde2009,Afshordi2008,Desjacques2010}, since the local PNG bispectrum peaks in squeezed triangles. A similar scale-dependent bias correction can be derived for any general nonlocal quadratic non-Gaussianity template \cite{Schmidt2010,Scoccimarro2011,Desjacques2011b,Schmidt2013}. 
  %by taking the squeezed limit that is relevant for bias on the scales considered, and introduce an additional operator $\Psi$ in the expansion, which is a nonlocal transformation of $\Phi$ and mediates the modulation of the local power spectrum in the small scale patch (see e.g. Ref.~\cite{Desjacques2016} for a review). For an arbitrary quadratic PNG template, 
  Then the bias expansion, linear in $\fnl$, is \cite{Assassi2015}:
  \begin{equation}\label{eq:deltaNG}
   \delta_h^{E,nG}(\bx,\tau)=b_{\Psi}^E(\tau)\Psi(\bq) +b_{\Psi\delta}^E(\tau)\Psi(\bq)\delta(\bx,\tau)+\veps_{\Psi}^E(\bx,\tau)\Psi(\bq)\,,
  \end{equation}
  where $\bq$ are the spatial coordinates in the Lagrangian frame, $\Psi$ is a nonlocal transformation of $\Phi$ and $\varepsilon_\Psi$ is its stochastic counterpart.
  
  By extending the peak-background split  argument (see e.g.~\cite{Desjacques2016} for a review) in the presence of PNG, we can derive the bias coefficients of the field $\Psi$. We consider a universal mass function $n_h(M,z)=(\mean{\rho}_m/M) f(\nu)|d\ln\nu/d\ln M|$, where the peak height is $\nu=\delta_c/\sigma_R(M,z)$, $\delta_c=1.686$ and $\sigma_R^2$ is the variance of the linear density field smoothed with a top-hat filter $W_R(k)$ of radius $R$. Then \cite{Desjacques2011a,Schmidt2013,Desjacques2016}:
 \begin{equation}\label{eq:bpsi}
     b_\Psi^E(M,z)=A\fnl\left[2\delta_cb_1^{L}+4\left(\frac{d\ln\sigma_{R,-\alpha}^2}{d\ln\sigma_{R}^2}-1\right)\right]\frac{\sigma_{R,-\alpha}^2}{\sigma_{R}^2},
  \end{equation}
  and \cite{Giannantonio2010,Karagiannis2018}
   \begin{equation}\label{eq:bpsidE}
   b_{\Psi\delta}^E(M,z)=2A\fnl\left[\delta_c\left(b_2^E+\frac{13}{21}(b_1^E-1)\right)
   +b_1^E\left(2\frac{d\ln\sigma_{R,-\alpha}^2}{d\ln\sigma_{R}^2}-3\right)+1\right]\frac{\sigma_{R,-\alpha}^2}{\sigma_{R}^2}\,,
  \end{equation}
  where $ \sigma_{R,n}^2=(2\pi)^{-3}\int d^3\bk\, k^n W_R(k)^2P^{\rm L}_m(k,z) $ and $b_1^{L}$ is  the Lagrangian bias (for a derivation of \esref{eq:bpsi}{eq:bpsidE}, see e.g. ~\cite{Karagiannis2018}). In the case of local PNG, where $\alpha=0$ and $A=1$, the above expressions reduce to the well-known results $b_\Psi^E\rightarrow b_\Phi^E=2\fnll\delta_cb_1^{\rm L}$ \cite{Dalal2008,Slosar2008,Giannantonio2010} and $b_{\Psi\delta}^E\rightarrow b_{\Phi\delta}^E=2\fnll[\delta_cb_2^E+(13\delta_c/21-1)(b_1^E-1)]$ \cite{Giannantonio2010,Baldauf2011,Sefusatti2012}. The equilateral PNG case is  $\alpha=2,A=3$, and for orthogonal PNG $\alpha=1,A=-3$ \cite{Schmidt2010,Giannantonio2012}.
  
  The presence of PNG will also affect the halo mass function \cite{LoVerde2007}, introducing  additional scale-independent corrections to the bias parameters~\cite{Desjacques2009,Sefusatti2012}. In this work, we take into account these corrections for the linear and quadratic bias coefficients, following \cite{Karagiannis2018}.

  %\section{The distribution of neutral hydrogen in the halo model}\label{sec:halo_bias}
  
  %A relation between the halo distribution and that of HI is necessary for  a realistic model of the bispectrum that will allow us to forecast the PNG measurements from future HI surveys as accurately as possible. 
  The general bias expansion results above can be applied to HI using the halo model framework \cite{Seljak2000,Peacock2000,Scoccimarro2000}, which describes how a tracer occupies the halo distribution. In this approach, HI is assigned only to halo hosts, with negligible contribution outside of them. The HI density is defined as \cite{Villaescusa2014,Castorina2016}: 
  \begin{equation}
      \rho_{\rm HI}(z)=\int d\ln M\, n_h(M,z) M_{\rm HI}(M,z) ,
  \end{equation}
  where $M_{\rm HI}$ is the average HI mass within the halo of total mass $M$ at redshift $z$. For $n_h(M,z)$ we use the fitting formula of  \cite{Tinker2008}, which is calibrated against N-body simulations.
  
  For $M_{\rm HI}$ {we use a halo} occupation distribution (HOD) approach \cite{Cooray2002} {and follow the model} of \cite{Castorina2016}:  
  \begin{equation}\label{eq:HOD}
    M_{\rm HI}(M,z)=C(z)(1-Y_p)\frac{\Omega_b}{\Omega_m}\,{\rm e}^{-M_{\rm min}(z)/M}\,M^{q(z)},
  \end{equation}
  where $C$ is a normalization constant, which cancels out in the HI bias,  $Y_p = 0.24$ is the helium fraction, $M_{\rm min}$ is the halo mass below which the amount of HI in halos is exponentially suppressed,  and $q$ controls the efficiency of processes generating or destroying HI inside  halos. 
 % $\Omega_{\rm HI}(z)=\rho_{\rm HI}(z)/\rho_c(0)$, where $\rho_c$ is the critical density. 
  The HOD of \eref{eq:HOD} is a power-law which agrees with the numerical results found in hydrodynamic simulations of~\cite{Villaescusa-Navarro:2015cca,Villaescusa-Navarro:2015zaa}. The exponential cut-off ensures that the amount of HI in low-mass halos is insignificant \cite{Pontzen:2008mx,Marin2010,Villaescusa2014}. 
  For the free parameters we choose ${q}=1$ and $M_{\rm min}=5\times 10^9 M_{\odot}/h$. 
 
 The HI bias parameters are 
  \begin{equation}\label{eq:bHI}
     b_{\rm HI}^i(z)=\frac{1}{\rho_{\rm HI}(z)}\int_0^\infty  d M\, n_h(M,z) b_i^h(M,z)M_{\rm HI}(M,z),
  \end{equation}
   where the index $i$ corresponds to the subscripts of the bias terms in \esref{eq:deltaG}{eq:deltaNG}. For $b_1^h$ we use the fitting function of~\cite{Tinker2010}, which is in good agreement with numerical results, for both low and high masses (see e.g.~\cite{Lazeyras2016}). For the higher-order bias coefficients ($b_2^h$, $b_3^h$, $b_4^h$) we use the mass function of \cite{Tinker2008} and the peak-background split in order to derive the analytic expressions (see e.g.~\cite{Karagiannis:2019jjx} for details). These analytic expressions have been tested against simulations in~\cite{Lazeyras2016}, which shows that the predicted $b_2^h$ deviates from the numerical results at low masses; given the HI low-mass suppression, we do not expect this to have a significant impact in our model. In~\cite{Lazeyras2016} it is shown that the {Tinker model \cite{Tinker2008} together with} the peak-background split prediction for $b_2^h$ is better than the standard results of~\cite{ShethTormen2001}; in addition, $b_3^h$ and $b_4^h$ agree with the simulation measurements, which justifies our preference for this approach.

\section{HI intensity mapping \rev{power spectrum and} bispectrum in redshift space }\label{sec:HIbisp_RSD}

{Observationally we determine the redshift, not the physical distance to a patch of the Universe and so we need to take into account} the effect of redshift space distortions (RSD)  \cite{Sargent:1977,Kaiser1987,Hamilton1998}, including the ``fingers of god'' (FOG) effect \cite{Jackson1972} in the non-perturbative regime. RSD can be modelled perturbatively \cite{Verde1998,Scoccimarro1998}, by generalising the SPT kernels to include RSD and bias expansions. The FOG effect is treated phenomenologically, by introducing an exponential damping factor $D_\text{FOG}$, which models the suppression of  clustering power in redshift space. In this work we stay mostly in the perturbative regime, slightly venturing into semi-nonlinear scales for low redshift. Therefore we consider a tree-level description for \rev{ both the power spectrum and} bispectrum, which is an adequate description for HI on the scales considered here  \cite{Gil-Marin:2014sta,Lazanu2015b,Hashimoto:2017klo,Chan2017,Oddo:2019run,Agarwal:2020lov,MoradinezhadDizgah2020}. We take into account the parameter shift due to neglecting higher-order 1-loop contributions at the Fisher matrix level, through the theoretical errors approach (see \sref{sec:fisher}). 

In the presence of a general PNG, the tree-level HI power spectrum and  bispectrum in redshift space are given by:
     \begin{align}
     P_{\rm HI}^s(\bk,z)&=T_b(z)^2\left[D_\text{FOG}^P(\bk,z)Z_1(\bk,z)^2P_m^{\rm L}(k,z)+P_{\veps}(z)\right]+P_{\rm N}(\bk,z) \label{eq:Pgs},\\ 
   B_{\rm HI}^s(\bk_1,\bk_2,\bk_3,z)&= T_b(z)^3\bigg\{D_\text{FOG}^B(\bk_1,\bk_2,\bk_3,z)\bigg[Z_1(\bk_1,z)Z_1(\bk_2,z)Z_1(\bk_3,z)B_{I}(k_1,k_2,k_3,z) \nonumber \\ 
   &+\Big\{2Z_1(\bk_1,z)Z_1(\bk_2,z)Z_2(\bk_1,\bk_2,z)P_m^{\rm L}(k_1,z)P_m^{\rm L}(k_2,z)+2~ \text{perm}\Big\}\bigg] \nonumber \\
   &+2P_{\veps\veps_{\delta}}(z)\Big[Z_1(\bk_1,z)P_m^{\rm L}(k_1,z)+2~ \text{perm}\Big]+B_{\veps}(z)\bigg\}. \label{eq:Bgs} 
  \end{align}
{In HI IM,} $P_{\rm N}$ is the instrumental noise (see \sref{sec:surveys}) and the background temperature is $T_b=0.072(1+z)^{2.6}H_0/H(z)~\mu$K \cite{StageII2018}. 
%with $E(z)=\sqrt{\Omega_m(1+z)^3+\Omega_k(1+z)+\Omega_\Lambda}$, for the standard dark-energy model (\ie $w_0=-1,\;w_\alpha=0$). 
The general non-Gaussian redshift kernels up to second order are \cite{Baldauf2011,Tellarini2016,Karagiannis2018}:
  \begin{align}
   &Z_1(\bk_i)=b_1+f\mu_i^2+\frac{b_{\Psi}k_i^{\alpha}}{M(k_i)}, \label{eq:Z1}\\
   &Z_2(\bk_i,\bk_j)=b_1F_2(\bk_i,\bk_j)+f\mu_{ij}^2G_2(\bk_i,\bk_j)+\frac{b_2}{2} +\frac{b_{s^2}}{2}S_2(\bk_i,\bk_j) 
   \nonumber \\ &
   +\frac{f\mu_{ij}k_{ij}}{2}\left[\frac{\mu_i}{k_i}Z_1(\bk_j)+\frac{\mu_j}{k_j}Z_1(\bk_i)\right]
%   \nonumber \\ &
   +\frac{\left[b_{\Psi\delta}-b_{\Psi}N_2(\bk_j,\bk_i)\right]k_i^{\alpha}}{2M(k_i)}+\frac{\left[b_{\Psi\delta}-b_{\Psi}N_2(\bk_i,\bk_j)\right]k_j^{\alpha}}{2M(k_j)}. \label{eq:Z2}
  \end{align}
where $f$ is the linear growth rate, $\mu_i=\hat\bk_i\cdot\hat{\bm{z}}$, with $\hat{\bm z}$ being the line-of-sight vector, $\mu_{ij}=(\mu_ik_i+\mu_jk_j)/k_{ij}$ and $k_{ij}^2=(\bk_i+\bk_j)^2$. Note that we have suppressed the redshift dependence for brevity. The kernels $F_2(\bk_i,\bk_j)$ and $G_2(\bk_i,\bk_j)$ are the second-order symmetric SPT kernels \cite{Bernardeau2002}, while $S_2(\bk_1,\bk_2) = (\hat\bk_1\cdot\hat\bk_2)^2-1/3$ is the  tidal kernel \cite{McDonald2009,Baldauf2012}. The kernel $N_2(\bk_1,\bk_2) = (\bk_1\cdot\bk_2)k_1^2$  encodes the coupling of the PNG potential to the Eulerian-to-Lagrangian displacement field  \cite{Giannantonio2010,Baldauf2011}. 
  
  The FOG damping factors are \cite{Peacock1994,Ballinger1996}
  \begin{align}
 D_\text{FOG}^P(\bk)&=\exp\big[-\big(k\mu\sigma_P\big)^2\big],\\
 D_\text{FOG}^B(\bk_1,\bk_2,\bk_3)&=\exp\big[-\big(k_1^2\mu_1^2+k_2^2\mu_2^2+k_3^2\mu_3^2\big)\sigma_B^2\big],
  \end{align}
  %$D_\text{FOG}^P(\bk)=\exp\big[-\big(k\mu\sigma_P\big)^2\big]$ and $D_\text{FOG}^B(\bk_1,\bk_2,\bk_3)=\exp\big[-\big(k_1^2\mu_1^2+k_2^2\mu_2^2+k_3^2\mu_3^2\big)\sigma_B^2\big]$  \cite{Peacock1994,Ballinger1996}, 
  where the damping parameters $\sigma_P$ and $\sigma_B$ have fiducial value equal to the linear  velocity dispersion $\sigma_\upsilon$. The redshift space bispectrum \eref{eq:Bgs} is characterized completely by five
variables: three to define the triangle shape (e.g. the
sides $k_1$, $k_2$, $k_3$) and  two to characterize the orientation of the  triangle relative to the line-of-sight. Here we follow the angle parametrization of \cite{Scoccimarro1999}, where the polar angle is $\omega=\cos^{-1}(\hat{\bk}_1\cdot\hat{\bm z})$ and the azimuthal angle is $\phi$. Then $\mu_1=\cos\omega=\hat{\bk}_1\cdot\hat{\bm z}$, $\mu_2=\mu_1\cos\theta_{12}+\sqrt{1-\mu_1^2}\sin\theta_{12}\sin\phi$ and $\mu_3=-(k_1/k_3)\mu_1-(k_2/k_3)\mu_2$, where  $\cos\theta_{12}=\hat{\bk}_1\cdot\hat{\bk}_2$. Then  $B_{\rm HI}^s(\bk_1,\bk_2,\bk_3,z)=B_{\rm HI}^s(k_1,k_2,k_3,\mu_1,\phi,z)$.

We keep only the PNG terms that are linear in $\fnl$, since higher order $\fnl$ terms will not contribute with the fiducial choice of $\fnl=0$. Note that, the scale-independent non-Gaussian corrections discussed in \sref{sec:halo_bias} are absorbed in the linear and quadratic bias parameters in the expressions above. In the case of equilateral PNG ($\alpha=2$), the $b_\Psi k^\alpha/M(k)$ term in the $Z_1$ kernel \eref{eq:Z1} becomes scale-independent on large scales and therefore degenerate with the linear bias. Towards small scales $k\sim1/R_*$, with $R_*$  the scale of halo formation, there is a scale dependence introduced by the transfer function in $M(k)$, which however, makes this PNG correction term degenerate with higher-order derivative terms \cite{Assassi2015}. Due to these degeneracies, on both large and small scales, this $\fnl$ scale-dependent correction term does not have constraining power for equilateral PNG. Hence it is excluded from the linear kernel expression of the tree-level bispectrum, as well as from the HI power spectrum, in order to avoid numerical contributions in the PNG signal that would in principle be inaccessible by an LSS survey. \rev{This effectively means that only the HI bispectrum can contribute to the equilateral PNG signal, in the case of a combined power spectrum and bispectrum analysis.}

  The stochastic terms in \eref{eq:Bgs}  approach their asymptotic constant values as $k\rightarrow 0$, which are those predicted by Poisson statistics (shot noise), and these will be chosen as their fiducial values. In the HI halo model formalism (\sref{sec:halo_bias}), the shot noise term is given by \cite{Castorina2016}
  \begin{equation} \label{eq:IM_shot}
      P_{\rm SN}(z)=\frac{1}{\mean{n}_{\rm eff}(z)}=\frac{1}{\rho^2_{\rm HI}(z)}\int dM\, n_h(M,z)M_{\rm HI}^2.
  \end{equation}
 The effective number density can in turn be used for the fiducial values of all the stochastic contributions \cite{Schmidt2015,Desjacques2016}:
  \begin{equation}\label{eq:poisson_fid}
   P_{\veps}\equiv P_{\rm SN}, \quad P_{\veps\veps_{\delta}}=\frac{b_1}{2\mean{n}_{\rm eff}}, \quad B_{\veps}=\frac{1}{\mean{n}_{\rm eff}^2}.
  \end{equation}

\section{HI intensity mapping surveys}\label{sec:surveys}

Radio telescopes can be set up to measure the 3D power spectrum of HI intensity in two distinctive ways: 
\begin{itemize}
\item as interferometers correlating the signals from all dishes or dipole stations and immediately outputting the Fourier transform of the sky -- interferometer (IF) mode;
\item as dishes providing separate maps of the sky,  added to reduce noise, with the final map Fourier transformed --  single-dish (SD) mode.
\end{itemize}

The noise power is dominated by instrumental noise, with a much smaller shot-noise contribution \cite{2011ApJ...740L..20G}. For completeness, we include shot noise, as in \eref{eq:IM_shot}. In IF mode, a Gaussian model of instrumental noise is given by \cite{Zaldarriaga2003b,Tegmark2008}
 \begin{equation}\label{eq:PSN}
  P^{\rm IF}_{\rm N}(\bk,{z})=T_{\rm sys}(z)^2\chi(z)^2\lambda(z)\frac{(1+z)}{H(z)}\left[\frac{\lambda(z)^2}{A_{\rm e}}\right]^2\frac{1}{N_{\rm pol}\,n_{\rm b}(\bm{u},z)\,t_{\rm{survey} }}\frac{S_{\rm area}}{\theta_{\rm FOV}(z)^2}\,.
 \end{equation}
Although the instrumental noise depends only on technical specifications and survey details, computing its power spectrum requires assuming a cosmology, via $H(z)$ and the comoving distance $\chi(z)$. {This arises from the fact that the instrumental noise has to be projected into a physical {`voxel'}.}
The survey sets the integration time $t_{\rm{survey} }$ and the sky area $S_{\rm area}$, and $\lambda(z)=\lambda_{21}(1+z)$ is the observed wavelength of the 21cm line. In the case of dishes, the field of view $\theta_{\rm FOV}$ depends on the size of the dish $D_{\rm dish}$:
\begin{equation} \label{eq:FOV}
\theta_{\rm FOV}(z)=1.22\,\frac{\lambda(z)}{D_{\rm dish}}\,.
\end{equation}
In the case of stations composed of dipoles, \eref{eq:FOV} holds with $D_{\rm dish} \to D_{\rm station}$. The antenna distribution determines the baseline density $n_{\rm b}$ in the image plane (the $\bm{u}$-plane). For simplicity, we assume it to be uniform, so that \cite{Bull2015}
\begin{equation}\label{eq:unibase}
n_{\rm b}(\bm{u},z) \approx { \frac{N_{\rm dish}^2}{ 2\pi\, u_{\rm max}(z)^2}}\,, \qquad u_{\rm max}(z)=\frac{D_{\rm max}}{ \lambda(z)} \,, 
\end{equation}
where $D_{\rm max}$ is the maximum baseline and $N_{\rm dish}$ is the number of dishes (or stations). In this approximation,
$P^{\rm IF}_{\rm N}(\bk,{z})=P^{\rm IF}_{\rm N}({z})$. We also assume dual polarizations per feed, $N_{\rm pol}=2$. The effective area $A_{\rm e}=3\eta D_{\rm dish}^2/(4\pi)$ depends on the efficiency $\eta$ and we take $\eta=1$. The system temperature $T_{\rm sys}$ is the receiver temperature $T_{\rm rx}$ plus sky temperature $T_{\rm gal}$.\\

\begin{table}[t]
 \centering
 \begin{tabular}{l|cc|cc|c|c}
  SD Survey & \multicolumn{2}{c|}{MeerKAT} & \multicolumn{2}{c|}{SKA1-MID$^{a}$} & HIRAX & PUMA ($\rm{Full}$)\\ \hline\hline
  & L Band & UHF Band & Band 1 & Band 2 && \\ \hline 
   redshift   &  0.1$^{b}-$0.58 & 0.4$-$1.45 & 0.35$-$3.05 & 0.1$^{b}-$0.49 & $0.75-2$ & $2-6$  \\
 $N_{\rm dish}$ & $64$ & $64$ & $197$ & $197$ & $1,024$ & $32,000$ \\
 $D_{\rm dish}$ [m] & $13.5$ & $13.5$ & $15$ & $15$ & $6$ & $6$ \\
 %$D_{\rm max}$ [km] & $\sim5^b$ & $\sim5^b$ & $\sim5$ & $\sim5$ & $0.25$ & $0.7$ \\
 $S_{\rm area}$ [$\rm{deg}^2$] & $4,000$ & $4,000$ & $20,000$ & $20,000$ & $15,000$ & $\sim20,000$ \\
 $t_{\rm survey}$ [hrs] & $4,000$ & $4,000$ & $10,000$ & $10,000$ & $28,000$ & $40,000$ 
 \end{tabular}
 \caption{Telescope and survey details, for single dish mode. Notes: (a) the 64 MeerKAT dishes included in SKA1-MID will keep their original specifications. For simplicity we neglect this difference and assume all dishes with SKA1-MID details (see \cite{Fonseca:2019qek} for an accurate treatment). (b) band covers redshift range $z=0-0.1$ which we neglect.}
 \label{table:survey_specs_SD}
\end{table}
\begin{table}[t]
 \centering
 \begin{tabular}{l|c|c}
  IF Survey & SKA1-LOW & SKA2-LOW\\ \hline\hline
   redshift   &  $3-5$ & $3-5$  \\
 $N_{\rm dish}$ & $224$ & $\sim7000$\\
 $D_{\rm dish}$ [m] & $40$ & $6$ \\
 $D_{\rm max}$ [km] & $1$ & $\sim1$\\
 $S_{\rm area}$ [$\rm{deg}^2$] & $5,000$ & $\sim21,000$\\
 $t_{\rm survey}$ [hrs] & $5,000$ & $10,000$
 \end{tabular}
 \caption{ As in Table \ref{table:survey_specs_SD}, for interferometer mode surveys. The frequency range allows for $z>5$ but we exclude this to avoid the complexities of incomplete reionisation at high $z$.}
 \label{table:survey_specs_ITF}
\end{table}

For SD mode, the instrumental noise is \cite{Santos2015}:
\begin{equation}\label{eq:Pnoise_SD}
 P^{\rm SD}_{\rm N}({z})=T_{\rm sys}(z)^2\chi(z)^2\lambda(z)\frac{(1+z)}{H(z)}\, \frac{S_{\rm area}}{2 N_{\rm dish}t_{\rm survey}} % \eta^2 N_{\rm feed} 
 \,,
\end{equation}
where we assume that the dishes have a single feed and that $\eta=1, N_{\rm pol}=2$. 
We can include the effects of the telescope beam in the noise power spectrum, by multiplying this expression by $\exp\big[C\big(k_\perp \chi \lambda/D_{\rm dish} \big)^2 \big]$, where $C$ is a constant \cite{Bull2015}. We follow the alternative, which is to impose a cut-off on $k_\perp$ -- see \eref{eq:kperSDcuts}.

We consider surveys in SD and IF modes, for which the constraints on PNG from the \rev{ combined power spectrum and} bispectrum {signal} have not previously been presented -- see \tref{table:survey_specs_SD} and \tref{table:survey_specs_ITF} for the details. This includes current, near-future and more futuristic surveys: 
MeerKAT\footnote{www.ska.ac.za/science-engineering/meerkat/} \cite{Santos:2017qgq}, an already-operational precursor for SKA1-MID\footnote{www.skatelescope.org} \cite{Bacon:2018dui} -- both telescopes in SD mode; SKA1-LOW \cite{Bacon:2018dui} and its futuristic upgrade SKA2-LOW \cite{Pourtsidou:2016ctq}, in IF mode;  
HIRAX \cite{Newburgh2016} and the futuristic PUMA \cite{PUMA_surv}, both in SD mode (in IF mode, their power spectrum and bispectrum constraints on PNG have been investigated in \cite{Karagiannis:2019jjx}).

Cosmological survey specifications for MeerKAT are from \cite{Santos:2017qgq}, for SKA1 from \cite{Bacon:2018dui}, for HIRAX from \cite{Newburgh2016,Karagiannis:2019jjx}, for PUMA from \cite{PUMA_surv,Karagiannis:2019jjx} and for SKA2 from \cite{Pourtsidou:2016ctq}. 
For HIRAX and PUMA, we assume the effective dish area is $D_{\rm eff}=\sqrt{\eta_a} D_{\rm dish}$, due to the non-uniform illumination of the primary, where $\eta_a=0.7$ is the aperture efficiency factor \cite{StageII2018}. Consequently, in all general expressions we apply $D_{\rm dish}\rightarrow D_{\rm eff}$ for HIRAX and PUMA. The system temperature for MeerKAT and SKA follows \cite{Bacon:2018dui}, 
%\footnote{\url{https://docs.google.com/document/d/1U-n9wJ91EzwbXo7vekn0K1I2V_lxY3AVQdsHs12WcTA/edit}}. 
and for HIRAX and PUMA we use \cite{StageII2018}. 

%In \tref{table:survey_specs_ITF} we summarize the details of the experiments we consider in ITF mode, while in \tref{table:survey_specs_SD} we summarize the SD mode experimental details. Note that we have made several simplifications. The MeerKAT's dishes will be integrated into SKA1 MID together with 133 15m dishes and fitted with band1 and band2. The final product is a mixture of both dish sizes, system temperatures, and bands. To avoid over-complications we take all the SKA1 MID to have the same specifications as listed in Tables \ref{table:survey_specs_ITF} and \ref{table:survey_specs_SD}. Also note that while we use all SKA1 MID dishes in SD mode, in ITF it is not advantageous to do so. In fact, the extra long-baseline dishes will only add information in very small nonlinear scales that we are not considering currently. Hence for the ITF we only consider the antennas in the core of the telescope. 

\section{Forecasting method}\label{sec:fisher}

% In the case of the galaxy power spectrum, the Fisher matrix is given by
%   \begin{equation}\label{eq:fisherPs}
%   F_{\alpha\beta}^{Ps}=\sum_{\mu_1}^{1}\sum_k \frac{\partial P_g^s}{\partial p_{\alpha}}\frac{\partial P_g^s}{\partial p_{\beta}}\frac{1}{\Delta P^2} \, ,
%  \end{equation}
  
%  \noindent while 

We predict the precision of the PNG amplitude measurement from the surveys considered in \sref{sec:surveys} by utilising the Fisher information matrix formalism. In order to derive the covariance matrix, we approximate the surface around the maximum peak of the likelihood distribution with a multivariate Gaussian. This is not generally true for a cosmological parameter, although it is a reasonable approximation near the peak. To improve upon this, one would need to sample the likelihood at various points in a multi-dimensional parameter space, which can be a very demanding process. The Fisher matrix formalism is much faster than parameter-space sampling, and delivers a reasonable approximation of the parameter uncertainties and correlations. 

 \begin{figure*}[t]
\centering
\resizebox{0.65\textwidth}{!}{\includegraphics{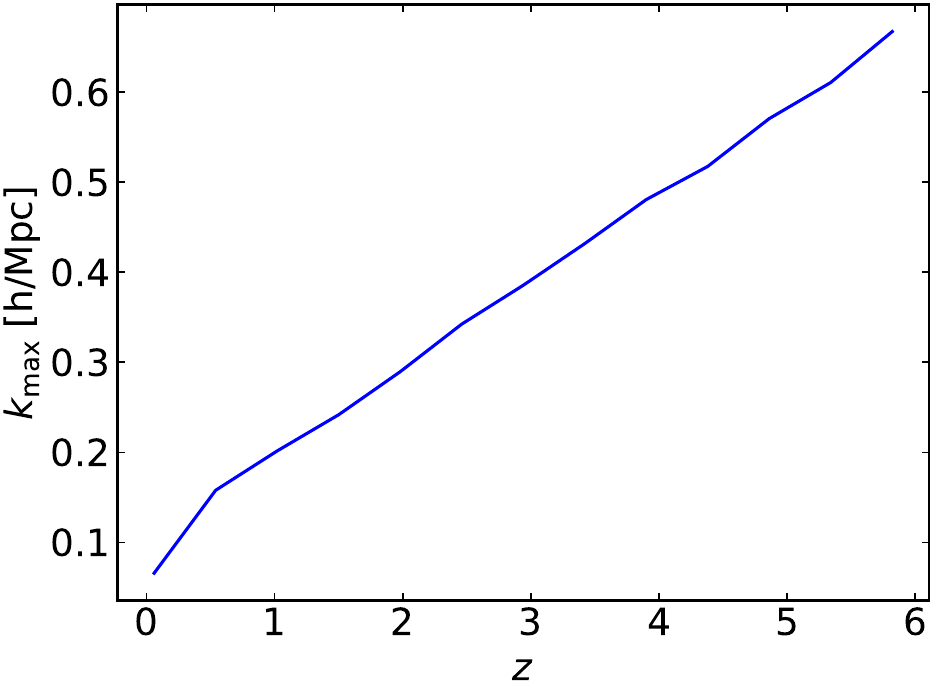}}
  \caption{The small scale cut-off, $k_{\rm max}(z)=0.75\,k_{\rm NL}(z)$, chosen here as a function of redshift.} 
\label{fig:kmax}
\end{figure*}

%Taylor expanding around the maximum likelihood value of the parameters $\boldsymbol{\theta}_0$, as derived from their estimator $\hat{\boldsymbol{\theta}}$, gives the Fisher information matrix
%\begin{equation}
 % F_{\alpha\beta}=-\left\langle\,\frac{\partial^2\ln L(\bx_0;\boldsymbol{\theta})}{\partial\theta_\alpha\partial\theta_\beta}\,\right\rangle_{\boldsymbol{\theta}=\boldsymbol{\theta}_0},
%\end{equation}
%where $L$ is the likelihood, $\bx_0$ is the dataset corresponding to the maximum likelihood parameters and $\boldsymbol{\theta}$ is the parameter vector defined in \eref{eq:params}. The inverse Fisher matrix $(F^{-1})_{\alpha\beta}$ yields an estimate of a parameter covariance with the smallest possible uncertainty on the measurement errors, so that the minimum error on a parameter is $\sigma(\theta_\alpha)=\sqrt{(F^{-1})_{\alpha\alpha}}$.

% For the HI IM bispectrum, the Fisher matrix at redshift $z$ is \cite{Scoccimarro2003,Agarwal:2020lov}
%   \begin{equation}\label{eq:fisherBs}
%   F_{\alpha\beta}^{B}(z)=\frac{1}{4\pi}\sum_{k_1,k_2,k_3}\int_{-1}^{1}d\mu_1\int_0^{2\pi}d\phi\,\frac{\partial B_{\rm HI}^s(\bk_i,z)}{\partial \theta_{\alpha}}\frac{\partial B_{\rm HI}^s(\bk_i,z)}{\partial \theta_{\beta}}\frac{1}{\Delta B^2(\bk_i,z)}\, ,
%   \end{equation}

\rev{For the HI IM power spectrum, the Fisher matrix at redshift $z$ is \cite{Tegmark1997}
\begin{equation}\label{eq:fisherPs}
    F_{\alpha\beta}^{P}(z)=\sum_k\int_{-1}^1 \frac{d\mu}{2} \frac{\partial P_{HI}^s(\bk,z)}{\partial \theta_{\alpha}}\frac{\partial P_{HI}^s(\bk,z)}{\partial \theta_{\beta}}\frac{1}{\Delta P^2(\bk,z)} \, ,
   \end{equation}
   while for the bispectrum}
   \begin{equation}\label{eq:fisherBs}
  F_{\alpha\beta}^{B}(z)=\frac{1}{4\pi}\sum_{k_1,k_2,k_3}\int_{-1}^{1}d\mu_1\int_0^{2\pi}d\phi\,\frac{\partial B_{\rm HI}^s(\bk_i,z)}{\partial \theta_{\alpha}}\frac{\partial B_{\rm HI}^s(\bk_i,z)}{\partial \theta_{\beta}}\frac{1}{\Delta B^2(\bk_i,z)}\, ,
  \end{equation}
 where $\theta_{\alpha}$ are the parameters, and the sum over triangles has $k_{\rm min}\le k_3\le k_2\le k_1 \le k_{\rm max}$. The bin size $\Delta k$ is taken to be the fundamental frequency of the survey, $k_{\rm f}=2\pi/L$, where for simplicity we approximate the survey volume as a cube, $L=V_{\rm s}^{1/3}$. The minimum value is $k_{\rm min}=k_{\rm f}$, which is the largest scale available to the survey, and the maximum value $k_{\rm max}$ corresponds to the smallest scale where the theoretical model is reliable. We follow \cite{Karagiannis:2019jjx} and set $k_{\rm max}(z)=0.75\,k_{\rm NL}(z)$, where $k_{\rm NL}$ is given by the one-dimensional velocity dispersion, i.e., $k_{\rm NL}(z)^{-2} = \int_0^\infty dk\, P^{\rm L}_m(k,z)/(6\pi^2)$. The choice of $k_{\rm max}$ confines the analysis within the perturbative regime, where the tree-level {description} offers a good agreement with numerical results \cite{Gil-Marin:2014sta,Lazanu2015b,Hashimoto:2017klo,Chan2017,Oddo:2019run}. We show $k_{\rm max}(z)$ in Figure \ref{fig:kmax}.
 We also take into account the error from excluding next-to-leading order corrections through the theoretical errors approach \cite{Baldauf2016}, which we discuss below. 
 
 The increase of $k_{\rm max}$ with redshift reflects the fact that the Universe is more linear at higher redshifts, so that the tree-level {modeling} holds up to higher values of $k_{\rm max}$, allowing us to include more modes and thus more triangles, for a stronger bispectrum signal and reduced cosmic variance. \rev{On the other hand, the bulk of the power spectrum signal originates from the largest accessible scales, due to the scale-dependence in the PNG bias correction.} Exploiting the azimuthal symmetry of $B_{\rm HI}^s$, which depends on $\phi$ only through $\sin\phi$, we can integrate over $\phi$ from $\pi/2$ to $3\pi/2$ and multiply the integral by 2, speeding up the numerical calculations significantly.
  
%  From the bispectrum estimator defined in Ref.~\cite{Scoccimarro1998}, we can derive the covariance of the estimator, as was done in Ref.~\cite{Sefusatti2006}. In the Fisher matrix analysis applied here we will 
 We only consider the diagonal part of the \rev{power spectrum and} bispectrum covariance; %i.e. the variance $\Delta B^2$ of \eref{eq:fisherBs}) neglecting all the cross-correlations between different triangles. Adopting the and in the
 in the Gaussian approximation the variance is \cite{Sefusatti2006,Sefusatti2007}:
 \begin{align}
 &\Delta P^2(\bk,z)=\frac{4\pi^2}{V_{\text{s}}(z)k^2\Delta k(z)}P_{HI}^s(\bk,z)^2\, , \label{eq:deltaP2} \\
 &\Delta B^2(\bk_i,z)=s_{123}\,\pi\, k_{\rm f}(z)^3\,\frac{P_{\rm HI}^s(\bk_1,z)\,P_{\rm HI}^s(\bk_2,z)\,P_{\rm HI}^s(\bk_3,z)}{k_1k_2k_3\,[\Delta k(z)]^3}\, ,\label{eq:deltaB2}
  \end{align}
  where $s_{123}=6,2,1$ for equilateral, isosceles and non-isosceles triangles respectively. In addition, for degenerate configurations, i.e. $k_i=k_j+k_m$, the \rev{bispectrum} variance should be multiplied by a factor of 2 \cite{Chan2017,Desjacques2016}. The off-diagonal terms of the covariance are related to higher than three-point correlators \cite{Sefusatti2006}, making the numerical implementation extremely tedious. 
  For the high-density samples, redshift range and large scales considered in this work, we do not expect the exclusion of the off-diagonal part of the covariance to have a significant impact on PNG forecasts \cite{Chan2017} (see also \cite{Karagiannis2018}). On the other hand, higher-order corrections to the (diagonal) variance could have a significant effect, not only for cosmological parameters \cite{Chan2017}, but also for PNG constraints \cite{Karagiannis2018}. The non-Gaussian contributions can be approximated by including perturbative corrections to the bispectrum variance, using the prescription of \cite{Chan2017}:
  \begin{equation}\label{eq:DB2_NL}
 \Delta B_\text{NL}^2(\bk_i)=\Delta B^2(\bk_i)+\frac{s_{123}\,\pi\, k_{\rm f}^3}{k_1k_2k_3\,(\Delta k)^3}\Big[P_{\rm HI}^s(\bk_1)\,P_{\rm HI}^s\,(\bk_2)\,P_{\rm HI}^\text{NL}(\bk_3)+2\perm \Big].
\end{equation}
We have omitted the $z$-dependence {for simplicity. The nonlinear power spectrum} $P_{\rm HI}^\text{NL}(\bk)$ is given by \eref{eq:Pgs} after replacing the linear power spectrum with its nonlinear correction: $P_m^{\rm L}(k)\rightarrow P^{\rm NL}_m(k)-P_m^{\rm L}(k)$, where $P^{\rm NL}_m$ is the nonlinear power spectrum from Halofit \cite{Smith2003,Takahashi2012}. 
 
%We need to identify the parameters that are considered free and will be marginalised over, in order to take into account degeneracies and cross-correlations between the various unknown parameters of the model. 
We assume that the cosmological parameters are known, since they can be measured with high precision by the CMB and the power spectrum \cite{Shaw:2013wza,Alonso:2014dhk,Switzer:2015ria,Olivari:2015tka,Olivari:2017bfv}. For the CMB primordial bispectrum,  degeneracies between cosmological parameters and $\fnl$ are small \cite{Liguori2008}, and we do not expect that cosmological errors to significantly impact the $\fnl$ constraints from the HI bispectrum. \rev{This was also pointed out, for both the galaxy power spectrum and bispectrum, in~\cite{Giannantonio2012,Moradinezhad2018,Bellomo:2020pnw}.}
% We assume that the linear HI bias is determined by power spectrum measurements. From \eref{eq:poisson_fid} if $b_1$ is known then $P_{\veps}$ is determined by $P_{\veps\veps_{\delta}}$. Therefore, the set of parameters that we consider is
%  \begin{equation}\label{eq:params}
%   \boldsymbol{\theta}=\big\{\fnl,b_2,b_{s^2},f,\sigma_{B};\,P_{\veps\veps_{\delta}},B_{\veps}\big\}.
%   \end{equation}

 \rev{The set of free parameters that we consider is
       \begin{equation}\label{eq:params}
    \mathbf{p}=\{\fnl,b_1,b_2,b_{s^2},P_{\veps},P_{\veps\veps_{\delta}},B_{\veps},f,\sigma_{\rm P},\sigma_{\rm B}\}.
   \end{equation}  
Forecasts from the combined power spectrum and bispectrum Fisher matrices, i.e. $F_{\alpha\beta}^{P+B}=F_{\alpha\beta}^{P}+F_{\alpha\beta}^{B}$, are considered here, where we neglect the cross-covariance between the two matrices. This would have minimal impact on our forecasts \cite{Chan2017,Yankelevich:2018uaz}. The stochastic bias contributions $P_{\veps\veps_{\delta}}$ and $B_{\veps}$ are considered as nuisance parameters and are marginalised over to acquire the Fisher sub-matrix for the parameters of interest, i.e., $\fnl,b_1,b_2,b_{s^2},f,\sigma_{\rm P},\sigma_{\rm B}$. In the Fisher sub-matrix we marginalise over the remaining free parameters for each redshift slice (cross-redshift correlations are assumed to be zero) \cite{Wang2006}. We then sum $F^{P+B}_{\fnl\fnl}(z_i)$ over the redshift range to derive the final forecasts on the $\fnl$ marginalised errors, presented in \sref{sec:results}.}

% The stochastic bias contributions $P_{\veps\veps_{\delta}}$ and $B_{\veps}$ are considered as nuisance parameters and are marginalised over to acquire the Fisher sub-matrix for the parameters of interest, i.e., $\fnl,b_1,b_2,b_{s^2},f,\sigma_{\rm P},\sigma_{\rm B}$. In the Fisher sub-matrix we marginalise over the remaining free parameters for each redshift slice (cross-redshift correlations are assumed to be zero) \cite{Wang2006}. We then sum $F^B_{\fnl\fnl}(z_i)$ over the redshift range to derive the final forecasts on the $\fnl$ marginalised errors, presented in \sref{sec:results}.

As noted before, ignoring next-to-leading order (1-loop) terms in \rev{the power spectrum and} bispectrum modeling can affect the forecasts. In a perturbative approach, each order has a limited range of validity and the higher-order contributions can become important approaching the nonlinear regime. The analysis here is mostly confined within the perturbative regime, where the tree-level  {description} gives accurate predictions, and we do not expect nonlinear uncertainties to significantly affect our forecasts. Nonetheless, in the Fisher matrix analysis, we incorporate the uncertainty of the theoretical model by following the approach of \cite{Baldauf2016}. In this formalism, theoretical errors are defined as the difference between the chosen perturbative order (i.e., tree-level) and the next higher-order (i.e., 1-loop). {An envelope ${E_i}$  \rev{(where $i$ is the index of the different momentum configurations, i.e. number of bins and triangles for the power
spectrum and bispectrum respectively)} is fitted, bounding these errors. The theoretical error covariance is $C_{ij}^e=E_i\rho_{ij}E_j$, where the correlation coefficient $\rho_{ij}$ is assumed to follow Gaussian statistics.} The correlation coefficient takes into account the correlations between the different momentum configurations, making the impact of the theoretical errors independent of the $k$-binning and the correlation length \cite{Baldauf2016}. \rev{The final covariance used in the Fisher matrix [Eqs. \eqref{eq:fisherPs} and \eqref{eq:fisherBs}] will be the sum of the error covariance $\bm{C}^e$ and the respective correlator variance [i.e.  \eref{eq:deltaP2} and \eref{eq:deltaB2}].}

  The envelopes are taken from \cite{Karagiannis2018}, which extends the approach of \cite{Baldauf2016} to include the theoretical uncertainties from excluding the 1-loop terms of the matter and of the local-in-matter bias expansions (i.e., $b_1,\;b_2,\;b_3$, etc.) in the \rev{power spectrum and} bispectrum (see \cite{Karagiannis2018} for further details).
  
  %, while performing an extensive study on the impact of theoretical errors on various parameters, including $\fnl$. These 1-loop bias terms have been shown to have an accurate description, by comparison with N-body simulations and galaxy catalogues (see e.g. refs~ \cite{Scoccimarro2001b,Feldman2001,Verde2002,Marin2013,GilMarin2014}). Additional bias terms in the 1-loop order of the bispectrum, under the effective field theory approach to perturbation theory \cite{Baumann2012,Carrasco2012} and the general bias expansion, have been presented \cite{Desjacques2016,Desjacques:2018pfv,Agarwal:2020lov} and could be potentially important. The extension the theoretical error formalism to include these terms is left for a future work. A similar approach to treat the theoretical uncertainties can be also found in Ref.~\cite{Audren2013}. Alternatively, in the spirit of Ref.~\cite{Baldauf2016}, an additional set of nuisance parameters can be added, by introducing a term that mimics the contribution of the most relevant 1-loop counter-terms, that need to be marginalised over and can be understood as theoretical errors, see for instance \cite{Sprenger:2018tdb,Obuljen:2017jiy,Agarwal:2020lov}. 
 
   \section{Observational window}\label{sec:obs_window}
   
%   The cosmological 21cm  signal is orders of magnitude fainter than the spectrally smooth foreground emission from astrophysical sources \cite{Shaw:2013wza,Shaw:2014khi,Pober:2014lva,Byrne:2018dkh}. This impacts the long wavelength fluctuations along the line-of-sight, so that small radial wavenumbers $k_\parallel=k\mu$ are lost  \cite{Jacobson:2003wv,Furlanetto:2006jb,Liu2011,Liu2012,Shaw:2013wza,Shaw:2014khi}. 

    The cosmological 21cm signal is orders of magnitude fainter than the foreground emission from astrophysical sources \cite{Shaw:2013wza,Shaw:2014khi,Pober:2014lva,Byrne:2018dkh}. The separation between the two is based on the spectrally smooth nature of the foregrounds. This means that only the long-wavelength fluctuations along the line of sight are affected \cite{Jacobson:2003wv,Furlanetto:2006jb,Chang2007,Liu2011,Liu2012,Shaw:2013wza,Shaw:2014khi}; \ie the small radial Fourier modes $k_\parallel\,(=k\mu)$ are contaminated.

  %The separation of the signal from the foreground emission imposes a great challenge for the analysis of 21-cm measurements. However, the nature of the production mechanism of the radio foregrounds, which is mainly composed of free-free and synchrotron emission from our galaxy and other unresolved sources, is intrinsically very spectrally smooth. This characteristic allows the separation from the cosmological signal, which varies along the line-of-sight due to the underlying density field, without significant losses up to some small value of $\kpar$ \cite{Liu2011,Liu2012,Shaw:2013wza,Shaw:2014khi}. The smaller the value of $\kpar$ we aim to recover the more difficult the separation becomes. The precise value, below which the recovery is impossible, is unknown, while a range of limits has been proposed throughout the literature (see e.g. refs~\cite{Liu2011,Liu2012,Shaw:2013wza,Shaw:2014khi,Pober:2014lva}). In Ref.~\cite{Shaw:2014khi} it is shown that a foreground cleaning can be achieved, making inaccessible only the modes that satisfy $\kpar<0.02\;\Mpc$. 
  
  Reconstruction techniques have been developed, to estimate long modes from the knowledge of short modes. In the context of HI intensity mapping, this has been applied to recover long radial modes lost to foreground cleaning
  \cite{Zhu:2016esh,Karacayli:2019iyd,Modi:2019hnu}. Using the forward model reconstruction framework \cite{Jasche2010,Kitaura2013,Wang:2014hia,Jasche:2014vpa,Shaw:2014khi,Wang:2016qbz,Seljak:2017rmr,Modi:2018cfi}, modes down to $k\simeq 0.01\;\Mpc$ can be almost perfectly recovered, despite the fact that none of these modes are contained in the data \cite{Modi:2019hnu}. Further developments of this technique could increase the range of recovery.
   
To avoid the region that is contaminated by foregrounds and where reconstruction of lost modes is not possible, we impose a cut on the line-of-sight wavelengths: 
  \begin{equation}\label{kparmin}
 \kpar \geq\kparmin \qquad\mbox{where}~~\kparmin=0.005\;\Mpc~\mbox{or}~ 0.01\;\Mpc \,.   \end{equation}
 Thus modes with $\kpar <\kparmin$ are excluded from the Fisher analysis. For the value of the cut-off we choose an optimistic case, $\kparmin=0.005\;\Mpc$, and a less optimistic one, $\kparmin=0.01\;\Mpc$.

The spectrally smooth foregrounds allow for a data-cleaning process which can be achieved without losing much of the cosmological information. However, in reality the interferometer's chromatic response will cause the foregrounds to leak into different modes transverse to the line of sight, introducing an additional non-smooth foreground component (e.g. \cite{Pober2015,Seo2015}). This effect in known in the literature as the foreground wedge \cite{Liu2011,Liu2012,Parsons2012,Pober2014,Seo2015,Pober2015,Seo2015} and is defined by  
 \begin{equation}\label{eq:kwedge}
k_{\parallel}<k_{\rm wedge}\,\kperp \,,   
 \end{equation} 
 where
 \begin{equation}\label{eq:wedge_prim}
  k_{\rm wedge}(z)=\frac{\chi(z)H(z)}{c(1+z)}\sin\big[ 0.61 N_w \,\theta_{\rm FOV}(z)\big].
 \end{equation}
 Here we will consider $N_w=1$. Note, that the foreground wedge is not a fundamental astrophysical limitation and with an excellent baseline-to-baseline calibration it can be removed \cite{Seo2015}. Such precise calibration is extremely hard at the moment, therefore %Nevertheless, here 
 we will exclude all modes that satisfy \eref{eq:kwedge}.
  
 The telescope beam should be taken into account in the analysis. Due to the high-frequency resolution of the IM experiments, the effective beam in the radial direction can be ignored, and only the response due to the finite angular resolution needs to be considered \cite{Bull2015}. The expression of the effective beam in the transverse direction depends on whether the instrument is in a SD or IF mode. 
  
 In the case of an interferometer, the effective beam is determined by the baseline number density distribution $n_{\rm b}(u)$ of the experiment. A simple approximation that is often used is a uniform number density  \eref{eq:unibase}, which we adopt here for the experiments in IF mode. This leads to a scale-independent noise power spectrum. Since such a distribution is unrealistic, sharp k-cuts should be introduced, limiting the range of the transverse scales to \cite{Bull2015}: 
\begin{equation}\label{eq:kperITFcuts}
k_{\bot, {\rm min}}^{\rm IF}=\frac{2\pi}{\chi \theta_{\rm FOV}}\,, \qquad
k_{\bot, {\rm max}}^{\rm IF}=\frac{2\pi D_{\rm max}}{\chi \lambda}\,.
\end{equation}
These originate from the fundamental limitations of an interferometer to probe scales larger/smaller than those that correspond to their minimum/maximum baseline. 

For an experiment in SD mode, the instrument's angular response is taken into account by multiplying the expression of the noise power spectrum \eref{eq:Pnoise_SD} by $W^2(\kperp,z)=\exp\big[-\big(k_\perp \chi \lambda/D_{\rm dish} \big)^2/(8\ln2) \big]$. This limits the accessible range of transverse modes to:  
\begin{equation}
 \label{eq:kperSDcuts}
k_{\bot, {\rm min}}^{\rm SD}=\frac{2\pi}{\sqrt{\chi^2S_{\rm area}}}\,, \qquad
k_{\bot, {\rm max}}^{\rm SD}=\frac{2\pi D_{\rm dish}}{\chi \lambda}\,.
 \end{equation}
Instead of using $W^2(\kperp,z)$ in the noise power spectrum expression for SD mode, we will simply exclude all $\kperp$ that are outside of the above scale range. In the literature it is common to treat the damping of scales from the beam and the survey by including a Fourier window function (see for example, \cite{Bull2015,Bernal:2019jdo}).  Here we took a more conservative approach by imposing sharp cuts.

\section{Results}\label{sec:results}

 \begin{figure*}[t]
\centering
\resizebox{\textwidth}{!}{\includegraphics{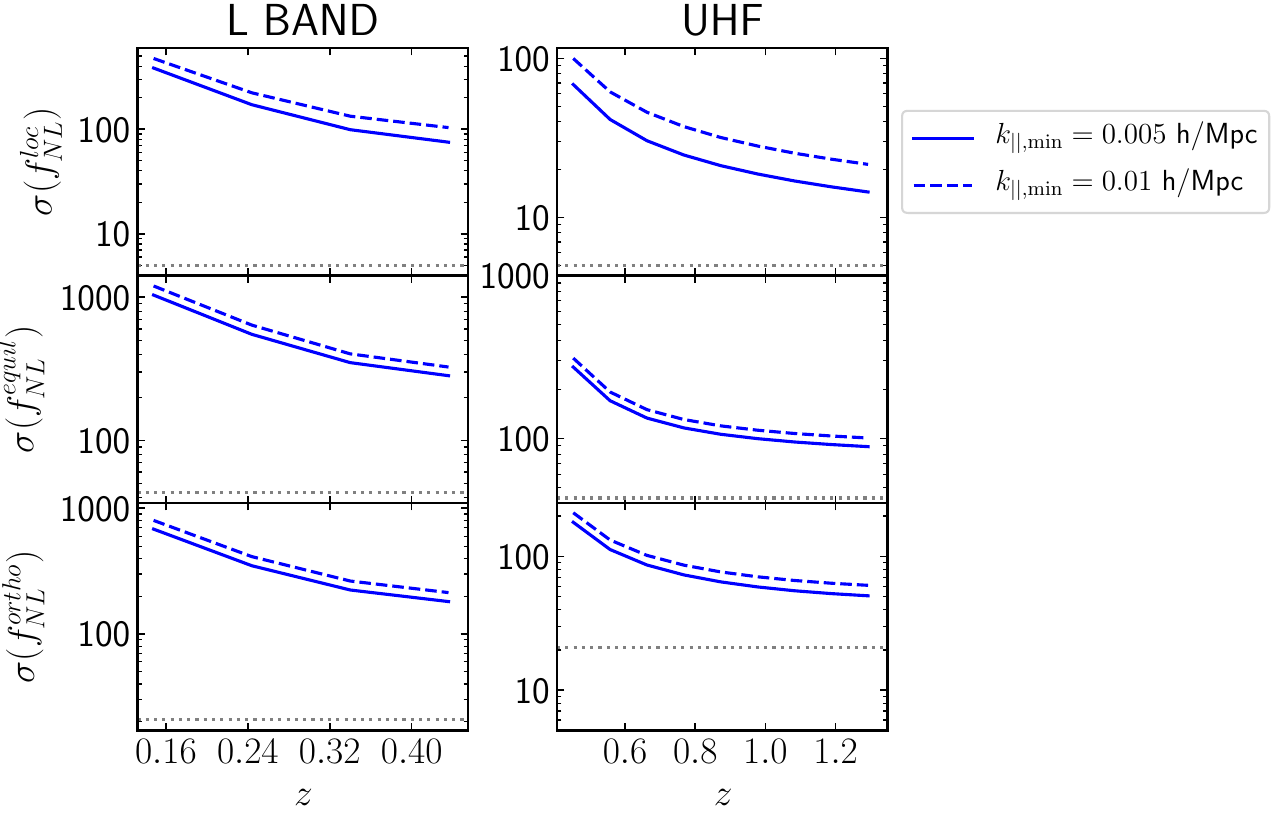}}
  \caption{Cumulative forecast error from \rev{the combined} HI IM \rev{power spectrum and} bispectrum \rev{signal} on $\fnll, \fnle, \fnlo$, for MeerKAT (L and UHF bands) in SD mode. Errors are shown for the two $\kparmin$ cuts in \eref{kparmin}. Grey dotted line denotes the Planck error {\cite{1905.05697}}. }
\label{fig:MK_cumul_sfnl}
\end{figure*}

The Fisher matrix forecast results from the \rev{combined HI power spectrum and bispectrum  signal} are presented in this section. The cumulative $1\sigma$ forecast error for the amplitude of the three PNG shapes, in the case of the SKA precursor MeerKAT, is presented as a function of redshift in \fref{fig:MK_cumul_sfnl}. The results for the two available MeerKAT bands are shown. \rev{The volumes probed by MeerKAT, as well as the scale limitations of the SD mode (see \sref{sec:obs_window}), restrict significantly the access to the large scales and therefore render the power spectrum contribution to the combined PNG signal minimal.} 

The UHF band corresponds to higher values of redshift, as well as to a more extended range, compared to the L-band, which leads to tighter constraints for all PNG cases. More precisely, almost an order of magnitude difference is observed between the forecasts originating from the two bands. The reason is that at higher redshifts, the observed physical volumes are larger and the accessible scales are probed with better resolution (\ie smaller fundamental frequency). In addition,  the clustering becomes more linear, so that the perturbative regime, where the tree-level description is adequate, can be pushed towards larger k-mode values (\ie increasing $k_{\rm max}$ with redshift). These two effects provide an increasing number of  triangles for the HI bispectrum,  hence boosting significantly the PNG signal towards increasing redshift values and improving the forecasts.

These conditions are responsible for the decreasing trend towards higher redshift values, up to a saturation point, where increasing redshifts do not contribute significantly to the cumulative signal. This is because the additional triangle configurations provided by the ever-increasing perturbative regime are excluded by the limitations of the MeerKAT in SD mode (see \sref{sec:obs_window}). In particular, for redshifts $z\gtrsim 0.4$ and $z\gtrsim 1.1$, in the case of the L and UHF band respectively, only a small contribution to the non-Gaussian signal is observed, for all three PNG types.

 \begin{figure*}[t]
\centering
\resizebox{\textwidth}{!}{\includegraphics{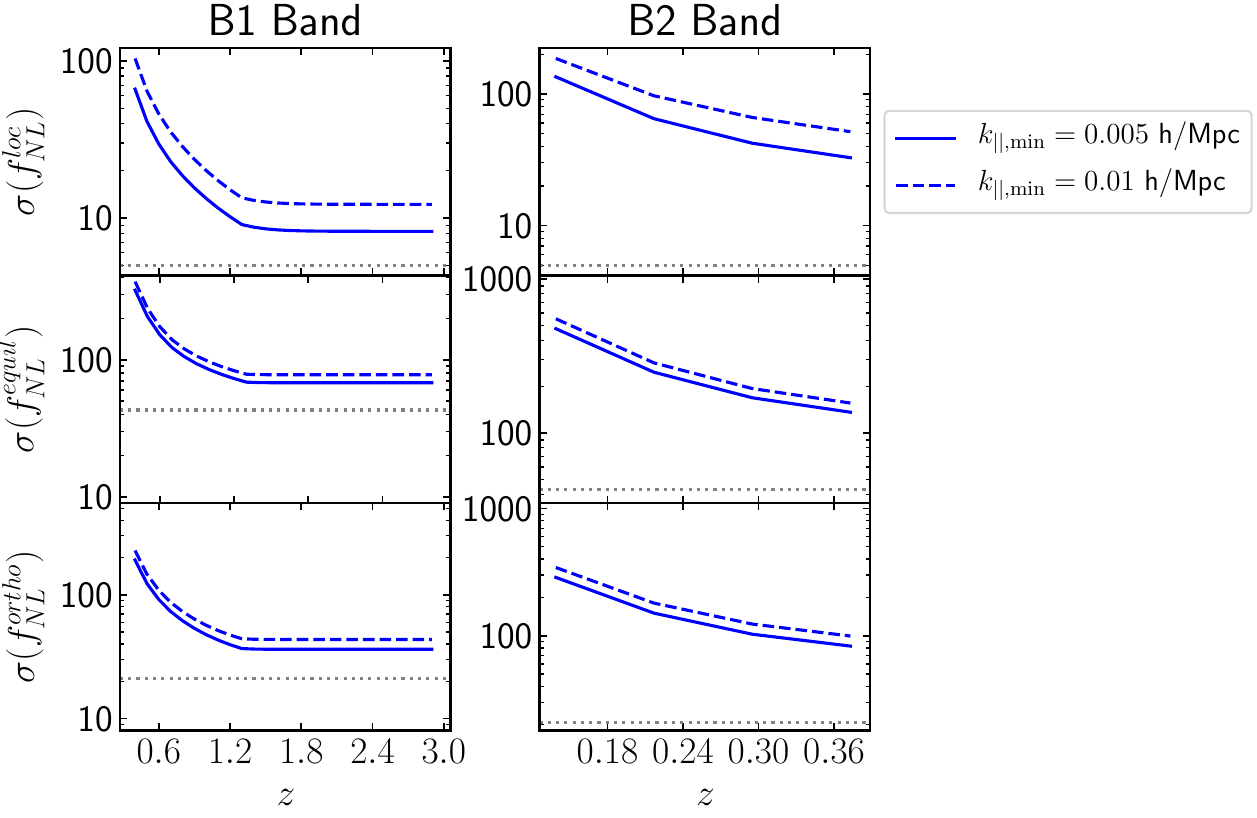}}
  \caption{As in \fref{fig:MK_cumul_sfnl}, for SKA1-MID in SD mode.} 
\label{fig:SKA_cumul_sfnl}
\end{figure*}

{If a larger number of large-scale modes is lost to foreground contamination, i.e. $\kparmin=0.01\;\Mpc$ (blue line), then all three PNG types are affected, with local PNG being the most affected.}  
A harder $\kparmin$ cut-off decreases the available signal inside the volume of a survey. Due to the instrumental and survey specifications, the UHF band is more affected, since a significant amount of large-scale modes, and hence squeezed configurations, get removed with the $\kparmin$ cut. This affects local PNG the most, \rev{since on the one hand the local PNG bispectrum peaks on the squeezed triangles and on the other the consequent limited access to the large scales, reduces significantly the PNG signal of the scale-dependent bias correction in the HI power spectrum contribution.} 
Equilateral PNG is affected the least, since the number of equilateral configurations, for which this template peaks, is not significantly reduced by the large-scale cut-offs. In fact, only a small number of the available equilateral triangles are excluded from each survey. \rev{Despite the fact that the power spectrum does not hold any constraining power on equilateral PNG (see \sref{sec:HIbisp_RSD}), adding the power spectrum Fisher matrix to the bispectrum one, improves the PNG constraints over the latter. This is due to the enhanced signal provided by the power spectrum on the redshift-dependent free parameters, e.g. $b_1$.}

Similar behaviour to the MeerKAT results can be seen in the case of the SKA1-MID experiment in SD mode  (\fref{fig:SKA_cumul_sfnl}). An improvement towards higher redshift, due to the larger accessible volumes, can be also observed for both bands considered. Band B1 results reach a saturation point around $z\sim 1.3$, while saturation in B2 is only in the last redshift bins. Despite the large sky area  and  extended redshift range of B1 band, the cumulative results over the whole redshift range are only marginally better than those provided by the MeerKAT high-redshift band (see \tref{table:forc_all}). The potential PNG signal in the \rev{power spectrum and} bispectrum, provided by the high redshift slices ($z>1.4$) of SKA1-MID B1 band, is excluded by the fundamental limitations of the SD mode, i.e. the telescope beam, and therefore there is no significant information to be added in the forecasts (see left panels of \fref{fig:SKA_cumul_sfnl}). \rev{Note, that due to the accessible scale range of the B1 band, the power spectrum provides equivalent $\fnl$ constraints to the bispectrum for local PNG, while for orthogonal PNG its contribution to the combined signal is marginal.}

On the other hand, constraints in the low-redshift band B2 of SKA1-MID  show a significant improvement ($\sim2$ times) over those from MeerKAT L-band (see Figs. \ref{fig:MK_cumul_sfnl},\ref{fig:SKA_cumul_sfnl} and \tref{table:forc_all}). This can be mainly attributed to the higher sky coverage, which offers a higher resolution in triangle-space, \rev{increasing the bispectrum signal}, since both surveys span roughly the same redshift range. Also, the instrumental noise of SKA1-MID is only marginally better than that of MeerKAT. \rev{Analogously to the previous cases, the power spectrum PNG constraints are insignificant compared to those provided by the bispectrum. However, as before, the power spectrum improves indirectly the $\fnl$ constraints, coming from the combined signal of the two correlators considered here, by reducing the correlation between the various redshift dependent parameters in $F_{\alpha\beta}^{P+B}$.}

Considering the most optimistic results ($\kparmin=0.005\;\Mpc$ cut-off) that originate from the cumulative signal over the total redshift range of SKA1-MID and MeerKAT (\tref{table:forc_all}), we can see that the constraints are far from being competitive compared to the Planck measurements \cite{Planck_PNG2016} (see also the grey dotted line in Figs. \ref{fig:MK_cumul_sfnl} and \ref{fig:SKA_cumul_sfnl}). Nonetheless, both MeerKAT UHF and SKA-MID B1 SD mode results are only few times worse than Planck, indicating that these HI IM data-sets can provide insightful preliminary tests on PNG and inflation.\\

The corresponding results in IF mode, for both surveys, are not shown. The specifications of MeerKAT and SKA-MID (see Tables \ref{table:survey_specs_ITF} and \ref{table:survey_specs_SD}) were not planned to function as a dedicated HI intensity mapping experiment. Therefore, they are not able to probe the corresponding scales that would make a bispectrum analysis viable in IF mode. In other words, IF mode excludes all the information available in these surveys.

 \begin{figure*}[t]
\centering
\resizebox{\textwidth}{!}{\includegraphics{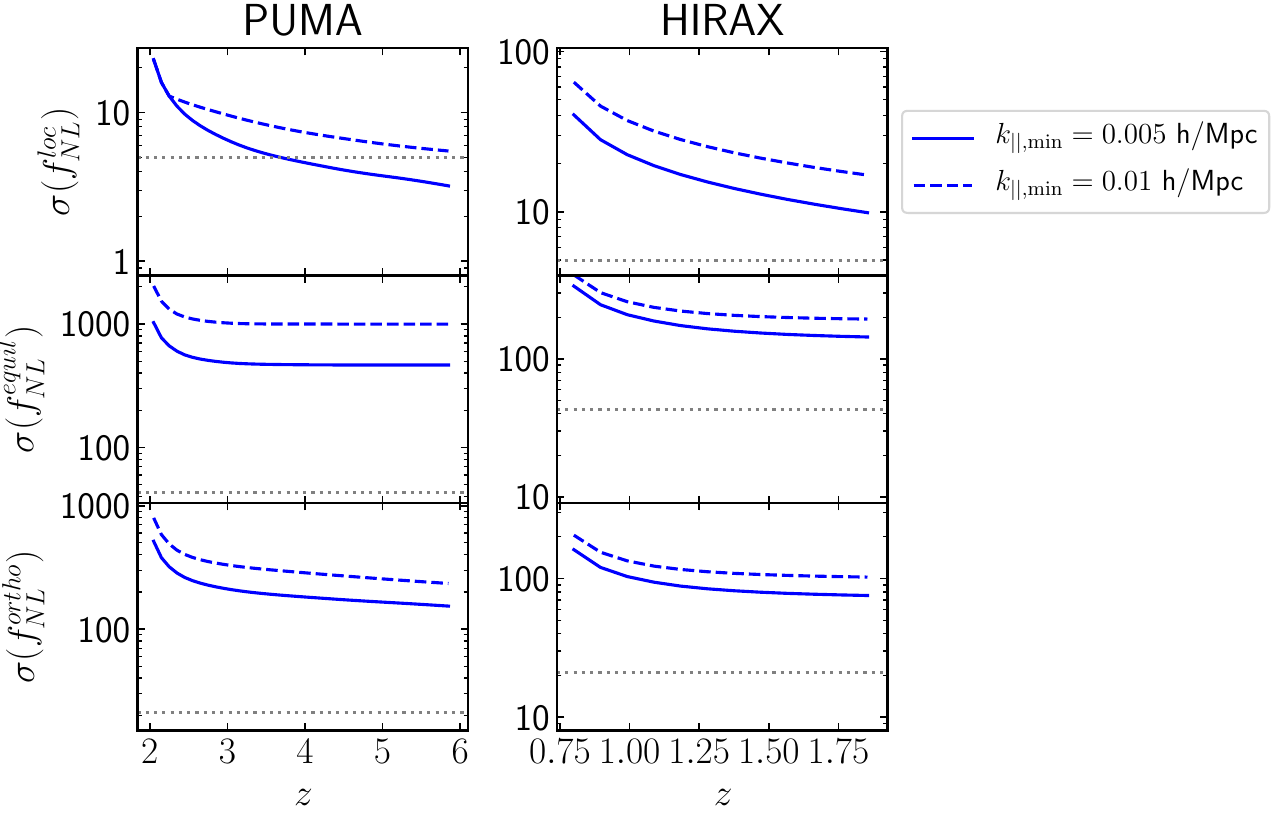}}
  \caption{As in \fref{fig:MK_cumul_sfnl}, for PUMA and HIRAX in SD mode.} 
\label{fig:PUMA_cumul_sfnl}
\end{figure*}

A bispectrum analysis is favoured by surveys that function as interferometers, especially for packed arrays which beat the instrumental noise, since they can probe large to intermediate scales, forming a significant number of triangles, that enhances the HI bispectrum signal \cite{Sefusatti2012}. This is particularly important for PNG constraints coming from IM experiments. Conversely, a power spectrum analysis is more suitable for IM surveys that were designed to function in SD mode. They offer a sufficient range and resolution over linear scales, for the scale-dependent bias correction (see \esref{eq:Pgs}{eq:Z1}) to provide compelling constraints on PNG. This is particularly true for the local type, which has been shown to be sufficiently constrained by the SKA in SD mode \cite{Camera2013,Fonseca:2015laa,2017MNRAS.466.2780F, Ballardini:2019wxj}. \rev{Nonetheless, due to the foregrounds and scale range probed by an experiment in SD mode, the bispectrum provides significantly tighter PNG constraints than the power spectrum for all the surveys discussed before, apart from SKA-MID B2 in the local PNG case. The latter is due to the fact that B2 offers a sufficient access to the large-scales, as well as the intermediate ones, providing an adequate amount of k-bins for the scale-dependent bias correction in the power spectrum to produce equivalent local PNG constraints to the bispectrum.}

The argument can be further supported by the results shown in the case of PUMA and HIRAX. Both are designed for IF mode. Nevertheless, they can still function as SD experiments (see \tref{table:survey_specs_SD}), where they can probe larger cosmological scales. The cumulative results from the \rev{ the combined HI power spectrum and bispectrum signal} are shown in \fref{fig:PUMA_cumul_sfnl}. An immediate observation is that the signal of PUMA saturates after the first couple of redshift slices ($z\sim2.7$), \rev{for equilateral and orthogonal PNG}. The same holds for HIRAX, where the saturation is evident beyond $z\sim 1.4$. For both surveys, the SD beam excludes a significant number of modes that are well inside the perturbative regime, which otherwise would have been accessible to the surveys at the higher redshifts. This is a common trait of all the surveys considered here that function in SD mode, \ie the bispectrum PNG signal is minimal beyond $z\gtrsim 1.4$, except for the full PUMA, which due to its large sky area coverage and redshift range, and its low instrumental noise, can provide sufficient PNG signal even at higher redshift values. 

\rev{This explains the functional behaviour over redshift for the cumulative error in the equilateral and orthogonal PNG cases, coming from the combined power spectrum and bispectrum signal, since the constraints on $\fnl$ originate mainly from the latter, while the contribution of the former remains minimal. On the other hand, for local PNG, the large volume of PUMA and HIRAX in SD mode provide  adequate access to the large-scale regime, which leads to the power spectrum constraints being the tightest, since the SD-mode scale limitations restrict  the squeezed configurations. The ever increasing linear regime, in conjunction with the increasing scale resolution (i.e. smaller $k_f$), results in improved local PNG power spectrum constraints towards high redshifts, justifying the observed redshift dependence of the error on $\fnll$ seen in the top panels of \fref{fig:PUMA_cumul_sfnl}.} 

Even with these improved specifications at hand, PUMA, as well as HIRAX, in SD mode do not offer competitive forecasts on \rev{equilateral and orthogonal} PNG types compared to Planck (\tref{table:forc_all}). \rev{In the case of local PNG, both surveys have the potential of providing competitive, or even improved in the case of PUMA, constraints relative to Planck.} In addition, comparing the cumulative errors \rev{on equilateral and orthogonal PNG} from both surveys with those provided by MeerKAT UHF and SKA1-MID B1, we observe that they are significantly less tight, e.g. PUMA SD-mode errors are $\sim 4$ times larger. This is because PUMA and HIRAX in SD mode do not utilise the notable bispectrum signal that lies in the higher redshift bins. \rev{On the other hand, for local PNG, PUMA is more promising, while HIRAX performs similarly to MeerKAT and SKA1-MID.}

The PUMA and HIRAX results complement the study of \cite{Karagiannis:2019jjx}, where both are considered in IF mode to forecast their power in constraining the three PNG types. A variety of observational effects, as well as different survey specifications, are studied in \cite{Karagiannis:2019jjx} to test their effect on the PNG signal, for both HI power spectrum and bispectrum. Comparing the results shown in \tref{table:forc_all} (SD mode results) with those presented in Table 2 of \cite{Karagiannis:2019jjx} (IF mode results), after considering the same $\kparmin$ sharp k-cuts (i.e. $\kparmin=0.01\;\Mpc$ for the foregrounds) as well as using $N_w=1$ for the foreground wedge in IF mode, we see that HIRAX in SD mode provides almost $\sim3-4$ times larger errors than in IF mode \rev{for the equilateral and orthogonal PNG cases, while for local the difference is reduced to $\sim 2$ times}. In neither of the two possible set-ups (i.e. SD and IF modes) can HIRAX  compete with the errors measured by Planck, although in IF mode they are only $\sim 2$ worse, as presented in \cite{Karagiannis:2019jjx}. 

In the case of PUMA, the change from SD to IF mode produces a significant difference in the PNG HI bispectrum forecasts. The SD constraints are far from being  competitive, \rev{with the exception of local PNG,} especially compared to smaller IM surveys, while for IF the PNG forecast errors can improve over those presented in Planck by a significant amount (see \cite{Karagiannis:2019jjx} for an extensive discussion). The large area covered by PUMA, leading to immense high-redshift volumes, offers a very high resolution in triangle space in the perturbative regime. This is utilised in the case of IF mode, in  contrast to SD, to provide very competitive constraints on PNG from the HI bispectrum. Furthermore, the tightly packed array of PUMA reduces significantly the instrumental noise, making available to the analysis a wide range of scales. \rev{On the other hand, the power spectrum takes advantage of the large-scale regime probed by PUMA in SD mode and the combined power spectrum and bispectrum signal provides competitive constraints, even with a minimal bispectrum contribution.} The comparison between SD and IF modes, in the cases of PUMA and HIRAX, further indicates that a HI bispectrum analysis, especially for PNG, is more suited to an IF than a SD experiment.

 \begin{figure*}[t]
\centering
\resizebox{\textwidth}{!}{\includegraphics{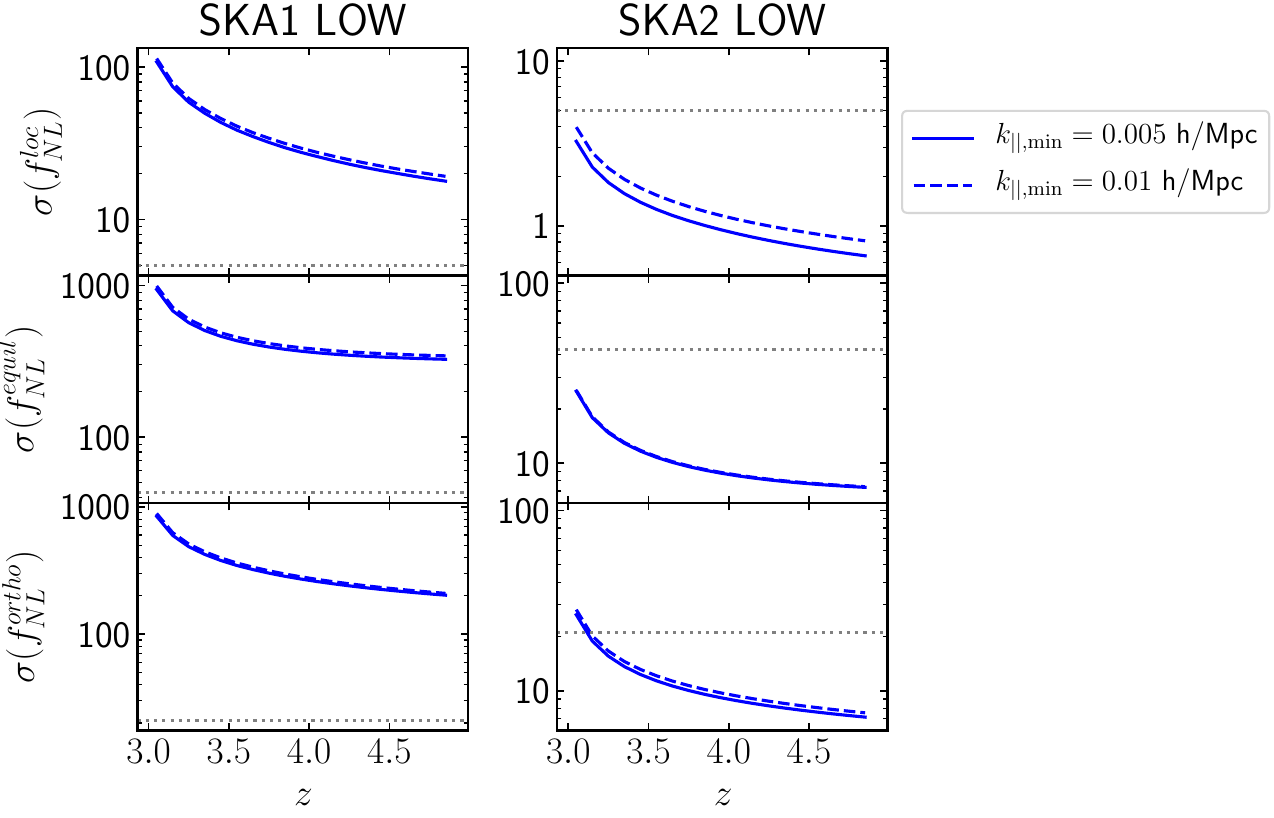}}
  \caption{As in \fref{fig:MK_cumul_sfnl}, for SKA1-LOW and SKA2-LOW in IF mode.} 
\label{fig:SKALOW_cumul_sfnl}
\end{figure*}

To further support this claim, we study PNG from the \rev{combined HI power spectrum and bispectrum signal} for SKA1-LOW and its improved futuristic version SKA2-LOW (\tref{table:survey_specs_ITF}) in IF mode. The cumulative errors on PNG are presented in \fref{fig:SKALOW_cumul_sfnl} as a function of redshift, while the errors from the cumulative signal over the whole redshift range are presented in \tref{table:forc_all} for all three PNG cases. The SKA1-LOW experiment was designed as an interferometer in order to study the epoch of reionisation. Nonetheless, the high redshifts provided, where the Universe is more linear, makes it ideal for a \rev{combined power spectrum and} bispectrum PNG analysis\footnote{The gravitational part of the HI bispectrum, originating from the late time evolution, will have a smaller amplitude at these redshifts.}. 

The cumulative results show a constant improvement over the whole redshift range, indicating that the scales from the growing perturbative regime, which are important for the HI bispectrum signal, are accessed in IF mode, especially those in the high redshift slices. \rev{Furthermore, the large scales probed by SKA-LOW provide enough PNG signal in order for the power spectrum to be the main contributor in the combined signal.} However, the low number of dishes, as well as the small sky area probed, increase the instrumental noise, thereby killing a significant amount of the available signal. These effects lead to non-competitive forecasts, in most cases even less than the low-redshift counterparts of MeerKAT and SKA1-MID.

Nevertheless, the improved version, SKA2-LOW, has the experimental settings, and in particular the large number of dishes, needed to deliver unprecedented constraints on PNG of all three types considered here, \rev{$\sigma(\fnll)\sim0.7$, $\sigma(\fnle)\sim13$ and $\sigma(\fnlo)\sim8$}. \rev{The bispectrum provides again here the bulk of the combined PNG signal, while the power spectrum in the case of local PNG, due to the extremely large volumes probed, offers a competitive amount of information with respect to the bispectrum, especially in the very high redshift slices.} In particular, SKA2-LOW improves over the Planck measurements already by utilising the bispectrum signal of the first couple of redshift slices (right panel of \fref{fig:SKALOW_cumul_sfnl}), in the case of the optimistic $\kparmin$ cuts. This is even true for the pessimistic sharp cut-off in the case of the equilateral PNG. This type of PNG is usually the least constrained by LSS surveys. Hence, the forecasts from SKA2-LOW on equilateral PNG indicate the great prospect that may lie within future IM interferometer surveys. The cumulative error shows a continuous decrease as a function of redshift, where it saturates only at the last high redshift bins. Excluding more large scales due to the foregrounds, still produces very competitive constraints (\tref{table:forc_all}), where local and orthogonal are affected the most due to the functional form of their templates (see \sref{sec:matter_PK_BK}).

\section{Discussion}\label{sec:discussion}

The goal of this paper was to study the \rev{ combined HI IM power spectrum and bispectrum} of current and planned experiments, in configurations not yet considered, in order to find the ideal set-up for \rev{such an} analysis,  in particular one targeting PNG. In the case of the MeerKAT, SKA-MID, HIRAX and PUMA experiments, we considered SD mode, while for SKA1-LOW and SKA2-LOW we considered IF mode. To determine the HI IM \rev{power spectrum and} bispectrum we reviewed the matter \rev{model}, the HI bias model, and the inclusion of RSD and PNG in the \rev{two correlators}. This work is also intended to complement other \rev{power spectrum and} bispectrum studies with LSS, using HI IM (see e.g. \cite{Watkinson:2017zbs,MoradinezhadDizgah:2018lac,Schmit:2018rtf,Karagiannis:2019jjx,Sarkar:2019ojl,Durrer:2020orn,Bharadwaj:2020wkc,Jolicoeur:2020eup}) and optical and radio continuum galaxy surveys (see e.g. \cite{Karagiannis2018}).

\begin{table*}[t]
\centering
\resizebox{\linewidth}{!}{
\begin{tabular}{c|cc|cc|c|c|c|c|}
\cline{2-9}
                & \multicolumn{2}{|c|}{MEERKAT (SD)} & \multicolumn{2}{c|}{SKA-MID (SD)} & \multicolumn{1}{c|}{PUMA (SD)} & \multicolumn{1}{c|}{HIRAX (SD)} & \multicolumn{1}{c|}{SKA1-LOW (IF)} & \multicolumn{1}{c|}{SKA2-LOW (IF)} \\ \hline
              \multicolumn{1}{|c|}{PNG Cases} & {L-BAND} & {UHF} & {BAND 1} & {BAND 2} & & & & \\ \hline              
\multicolumn{1}{|c|}{Local}   & 77 (105) & 15 (22) & 8 (12) & 33 (52) & 3 (6) & 10 (17) & 18 (19) & 0.7 (0.8)        \\
\multicolumn{1}{|c|}{Equilateral}  & 494 (578) & 141 (166) & 52 (60) & 232 (273) & 553 (1325) & 189 (274) & 607 (653) & 12.5 (12.8) \\
\multicolumn{1}{|c|}{Orthogonal}  & 234 (288) & 63 (80) & 47 (60) & 105 (134) & 154 (236) & 95 (139) & 215 (221) & 7.6 (8)      \\ \hline
\end{tabular}
}
\caption{Bispectrum forecasts for 1$\sigma$ marginalised errors on $\fnll, \fnle, \fnlo$, from the cumulative signal over the total redshift range. A foreground-imposed cut $\kparmin=0.005\; \Mpc$ is used. Errors in parenthesis correspond to a less optimistic $\kparmin=0.01\; \Mpc$. }
\label{table:forc_all}
\end{table*}

We started by analysing the prospects of the SKA precursor MeerKAT, as it is already operational. In \fref{fig:MK_cumul_sfnl} we showed the forecasted cumulative errors for the three $\fnl$ parameters as a function of redshift for the two MeerKAT available bands. An immediate observation is that the UHF band performs much better than the L-band. This comes as no surprise, because towards higher redshifts the observed physical volume increases, hence more triangles, \rev{as well as large-scale modes,} from which to extract information. By the same token all curves in \fref{fig:MK_cumul_sfnl} decrease with the increase of redshift, up to signal saturation. Also, as expected, including information from larger scales, i.e., having a lower $k_{\rm min}$ (red line), improves the forecasted errors. Although the MeerKAT is very far from providing a competitive error bar on the  \rev{ combined HI IM power spectrum and bispectrum signal} (especially in comparison with the latest Planck results \cite{1905.05697}, shown as the grey dashed line) it will provide an insightful data-set for preliminary tests. The cumulative constraints from all redshift bins, in the case of the pessimistic foreground cuts, are only few times worse than Planck (see \autoref{table:forc_all}).

SKA1-MID provided improved results relative to  MeerKAT (see \fref{fig:SKA_cumul_sfnl}). Such improvements mainly come from the larger sky area covered by the surveys, as the instrumental noise of  SKA1-MID  is only marginally better than  MeerKAT's. Both SKA1-MID bands follow the same trends as the MeerKAT ones, i.e.,  higher redshift bins contain a larger volume,  providing more available triangles \rev{and $k$-bins located in the large-scale regime} from which to extract information. Despite the improvement per redshift bin, the cumulative result (see \autoref{table:forc_all}) from the high-redshift SKA-MID survey is only marginally better than that from the  MeerKAT survey. On the other hand, for the low-redshift bands, the increased volume has a considerable impact on the forecasted error (by a factor $\sim2$).

Both MeerKAT and SKA-MID were not planned as dedicated HI intensity mapping facilities. Nevertheless, they  can be effective for the HI IM power spectrum. On the other hand, both HIRAX and PUMA were designed as dedicated experiments to extract the HI power spectrum in IF mode. Despite being designed as dish interferometers, they can still work as SD-mode experiments and therefore probe large cosmological scales: results are shown in \fref{fig:PUMA_cumul_sfnl}. These constraints are far behind the IF mode constraints presented in \cite{Karagiannis:2019jjx}. 

Finally, we studied the \rev{combined HI IM power spectrum and bispectrum} with SKA-LOW, which is designed as an interferometer to do HI IM of the epoch of reionisation. Nevertheless, its frequency range covers  redshifts  down to $z=3$ and so it can be used in IF mode for post-reionisation constraints on PNG. Our results are shown in \fref{fig:SKALOW_cumul_sfnl},  and we concluded that while SKA1-LOW will provide results similar to its low-redshift MID counterpart, SKA2-LOW will have enough information to hit the Planck level in single redshift bins, with cumulative constraints at unprecedented levels. 
Despite  attaining $\sigma(\fnll)\sim1$,  SKA2-LOW  is very futuristic. 

 In \autoref{table:forc_all} we summarised the results from all the experimental settings considered in this paper. While the local shape is the easiest to measure in all experiments, the equilateral shape is always harder. \rev{In the case of the bispectrum,} an explanation is due to the number of triangles available in a finite volume as well as the strength (or weakness) of the signal. The number of equilateral triangles inside a fixed scale range is the same as the number of individual k-bins, while there are more squeezed configurations that can be formed inside the same range. The number of the latter increases significantly if enough large scales are accessible to the survey. Furthermore, the equilateral PNG template peaks on the same configurations as the gravitational part of the bispectrum (equilateral triangles). The latter has a few orders of magnitude larger amplitude, making the disentanglement of the two a tedious process. \rev{Note, that for equilateral PNG, the power spectrum has no constraining power on $\fnle$ (see \autoref{sec:HIbisp_RSD}), making the bispectrum the sole contributor to the PNG signal.} Therefore, equilateral PNG is the least constrained shape from LSS surveys (see \cite{Karagiannis2018} for a discussion). More importantly, despite the ever-larger volumes of the surveys, the cuts on $\kpar$ from foregrounds in the signal impose  limitations on what one can do with the observed signal.  We imposed two cuts at $\kparmin=0.005\; \Mpc$ and a less optimistic $\kparmin=0.01\; \Mpc$. Fundamentally we know that for SD mode, the inclusion of larger scales improves the measurements substantially, especially for local PNG in the case of the power spectrum. This is also true in the case of the HI bispectrum, as it is shown in \tref{table:forc_all}. Hence it is crucial to obtain the signal from the larger scales instead of discarding it altogether. 
 
 One can try to disentangle the HI signal from the foreground signal using clever statistical methods. Indeed, a substantial amount of work has been done at the HI power spectrum level with blind foreground subtraction methods (see e.g. \cite{Alonso:2014dhk} for an extensive study of different foreground subtraction methods and \cite{Carucci:2020enz,Cunnington:2020wdu} for more recent studies). In general, blind foreground methods perform very well on small scales but do not recover well the large scales which are needed to improve bispectrum measurements. In practice, one would need to fit transfer functions based on simulations to fully reconstruct them. Alternatively one can look to reconstruct the large scales from other available scales using quadratic estimators. This has been proposed recently in \cite{Li:2020uug} and further explored in \cite{Li:2020luq} were RSD and bias models where included in their original work. This opens a new window to reconstruct large scales in HI intensity mapping that we will pursue in the future.

Our results indicate that a HI bispectrum analysis, and in particular a PNG-oriented one, is better done using an IF experiment with a packed array, i.e. high-density dish distribution. Although more and more dishes help to improve the noise, the two configurations considered provide information in disjoint scale ranges. While the size of the dish imposes a minimum accessible scale in SD modes, in IF mode it becomes the largest accessible scale. Hence, for bispectrum studies, one should choose a survey that includes the scales containing the bulk of the information, which are smaller but still linear scales. These are better probed by IF mode. In addition, IF surveys can be further improved by a large sky coverage and high redshift reach.  \rev{For the power spectrum, the bulk of the PNG signal, especially for the local case, originates from the large-scale regime, due to the scale-dependent bias correction. This makes SD-mode experiments ideal for a power spectrum analysis. Nonetheless, we show that in the case of small-volume surveys in SD mode, the bispectrum can still outperform power spectrum constraints. In the case where the large-scale regime is probed with sufficient resolution, the power spectrum provides the tightest constraints, in particular for the local PNG. In this case, for surveys that additionally offer access to scales towards the end of the linear regime, the power spectrum and bispectrum provide equivalent constraints and a combined analysis would optimally utilise the available PNG signal.} Proposed surveys that fit these criteria, like SKA2-LOW and PUMA (Full) (see \cite{Karagiannis:2019jjx}), make ideal candidates for putting tight constraints on the amplitude of PNG, including the elusive equilateral type, and shedding light on the inflationary epoch of the primordial Universe.

\bigskip

\section*{Acknowledgments}
DK and RM are supported by  the South African Radio Astronomy Observatory (SARAO) and the National Research Foundation (Grant No. 75415). 
JF was supported by the University of Padova under the STARS Grants programme {\em CoGITO: Cosmology beyond Gaussianity, Inference, Theory and Observations} and by the UK Science \& Technology Facilities Council (STFC) Consolidated Grant ST/P000592/1. JF also thanks the University of the Western Cape for supporting a visit during which parts of this work were developed.
 RM is also supported by the UK STFC Consolidated Grant ST/S000550/1. 
SC acknowledges support from the `Departments of Excellence 2018-2022' Grant (L.\ 232/2016) awarded by the Italian Ministry of University and Research (\textsc{mur}). SC is funded by \textsc{mur} through the Rita Levi Montalcini project `\textsc{prometheus} -- Probing and Relating Observables with Multi-wavelength Experiments To Help Enlightening the Universe's Structure'.



%% References



%% New version of the num-names style
\bibliographystyle{elsarticle-num-names}
\bibliography{bibliography.bib}

\end{document}